\documentclass[8.5pt,twoside,twocolumn]{article} 
\usepackage[super,sort&compress,comma]{natbib} 
\usepackage{times,mathptmx}

\usepackage{graphicx} 
\usepackage{lastpage}
\usepackage[format=plain,singlelinecheck=false,font=small,labelfont=bf,labelsep=space]{caption} 
\usepackage{fancyhdr}
\usepackage[usenames, dvipsnames]{color}
\usepackage{setspace}
\usepackage{amsmath,amssymb}
\usepackage{amsfonts}
\usepackage{bm}
\usepackage{multicol}

\newcommand{\mlmin}{\,\mathrm{ml/min}}

\newif\ifShapeComparison
\ShapeComparisontrue

\begin{document}

\thispagestyle{plain}
\fancypagestyle{plain}{
\fancyhead[L]{\includegraphics[height=8pt]{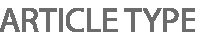}}
\fancyhead[C]{\hspace{-1cm}\includegraphics[height=20pt]{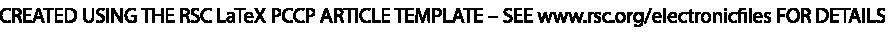}}
\fancyhead[R]{\includegraphics[height=10pt]{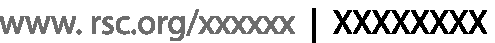}\vspace{-0.2cm}}
\renewcommand{\headrulewidth}{1pt}}
\renewcommand{\thefootnote}{\fnsymbol{footnote}}
\renewcommand\footnoterule{\vspace*{1pt}%
\hrule width 3.4in height 0.4pt \vspace*{5pt}} 
\setcounter{secnumdepth}{5}

\onehalfspacing

\makeatletter 
\def\subsubsection{\@startsection{subsubsection}{3}{10pt}{-1.25ex plus -1ex minus -.1ex}{0ex plus 0ex}{\normalsize\bf}} 
\def\paragraph{\@startsection{paragraph}{4}{10pt}{-1.25ex plus -1ex minus -.1ex}{0ex plus 0ex}{\normalsize\textit}} 
\renewcommand\@biblabel[1]{#1}            
\renewcommand\@makefntext[1]%
{\noindent\makebox[0pt][r]{\@thefnmark\,}#1}
\makeatother 
\renewcommand{\figurename}{\small{Fig.}~}

\fancyfoot{}
\fancyfoot[LO,RE]{\vspace{-7pt}\includegraphics[height=9pt]{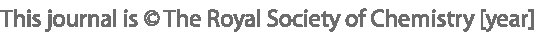}}
\fancyfoot[CO]{\vspace{-7.2pt}\hspace{12.2cm}\includegraphics{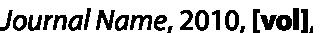}}
\fancyfoot[CE]{\vspace{-7.5pt}\hspace{-13.5cm}\includegraphics{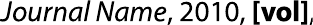}}
\fancyfoot[RO]{\footnotesize{\sffamily{1--\pageref{LastPage} ~\textbar  \hspace{2pt}\thepage}}}
\fancyfoot[LE]{\footnotesize{\sffamily{\thepage~\textbar\hspace{3.45cm} 1--\pageref{LastPage}}}}
\fancyhead{}
\renewcommand{\headrulewidth}{1pt} 
\renewcommand{\footrulewidth}{1pt}
\setlength{\arrayrulewidth}{1pt}
\setlength{\columnsep}{6.5mm}
\setlength\bibsep{1pt}

\twocolumn[
  \begin{@twocolumnfalse}
\noindent\LARGE{\textbf{Bubble propagation on a rail: a concept for sorting bubbles by size$^\dag$}}
\vspace{0.6cm}

\noindent\large{\textbf{Andr\'es Franco-G\'omez\textit{$^{a}$}, Alice B. Thompson\textit{$^{b}$}, Andrew L. Hazel\textit{$^{b}$}, and Anne Juel$^{\ast}$\textit{$^{a}$}}}\vspace{0.5cm}

\noindent\textit{\small{\textbf{Received Xth XXXXXXXXXX 20XX, Accepted Xth XXXXXXXXX 20XX\newline
First published on the web Xth XXXXXXXXXX 200X}}}

\noindent \textbf{\small{DOI: 10.1039/b000000x}}
\vspace{0.6cm}

\noindent \normalsize{We demonstrate experimentally that the introduction of a rail,
  a small height constriction, within the cross-section of a
  rectangular channel could be used as a robust passive sorting
  device in two-phase fluid flows. Single air bubbles carried within silicone oil are generally transported on one side of the rail. However, for flow rates marginally larger than a critical value, a narrow band of bubble sizes can propagate (stably) over the rail, while bubbles of other sizes segregate to the side of the rail. The width of this band of bubble sizes increases with flow rate and the size of the most stable bubble can be tuned by varying the rail width.
  We present a complementary theoretical analysis
  based on a depth-averaged theory, which is in qualitative agreement
  with the experiments. The theoretical study reveals
  that the mechanism relies on a non-trivial interaction between capillary
  and viscous forces that is fully dynamic, rather than being a
  simple modification of capillary static solutions.}
\vspace{0.5cm}
 \end{@twocolumnfalse}
  ]

\footnotetext{\dag~Electronic Supplementary Information (ESI) available: [details of any supplementary information available should be included here]. See DOI: 10.1039/b000000x/}


\footnotetext{\textit{$^{a}$~Manchester Centre for Nonlinear Dynamics \& School of Physics \& Astronomy, The University of Manchester, Manchester M13 9PL, UK. Tel: 0161 2754071; E-mail: anne.juel@manchester.ac.uk}}
\footnotetext{\textit{$^{b}$~Manchester Centre for Nonlinear Dynamics \& School of Mathematics, The University of Manchester, Manchester M13 9PL, UK.}}




\section{Introduction}\label{intro}

 The development of methods for sorting bubbles, droplets or particles
 within a suspending fluid is of fundamental importance
 to applications in biology, chemistry and
 industry. Consequently, a vast array of techniques for the active sorting
of suspended components in microfluidic devices have been
developed\cite{XiEtAl2017}, often with particular applications in
flow cytometry\cite{PiyasenaGraves2014}.
Active sorting methods rely on inducing
local motion of the suspended components
via external forcing, \textsl{e.~g.}
electrical, magnetic and acoustic forces or external pressure; and
typically require additional detection methods
to ensure that the external forcing is actuated at the appropriate point
in time. 
In contrast, passive approaches rely on harnessing the local
motions of the suspended objects
that arise through pure hydrodynamic interactions
between the flow, the suspended object itself and
the bounding geometry\cite{LealReview1980}; hence,
the presence of the particles does
not need to be detected. Thus passive sorting methods are simpler,
and potentially more robust and energy efficient
than their active counterparts.

 For flows within channels and tubes, 
hydrodynamic interactions can lead to migration of particles, bubbles and
droplets normal to the predominant direction of flow and
their accumulation at particular locations within
the channel cross-section\cite{Stan_etal2011}. Once such
spatial localisation has been achieved,
it is straightforward to use geometric
separators to collect the suspended objects with the desired properties.
Applications of passive methods include segregation of blood
components\cite{DiCarlo_etal2007}, trapping of micro-organisms in
water\cite{Kim_etal2014}, and purification of emulsions or
colloids.

The small scale of microfluidic devices means that
inertial migration mechanisms\cite{CoxMasonReview1971,Amini_etal2014}
do not necessarily operate effectively, although recent work has shown
that inertial microfluidics is feasible\cite{DiCarlo2009}. In
this paper we shall concentrate on hydrodynamic mechanisms that
operate on suspended gas bubbles and
rely on an interplay between surface tension and viscous forces.
The relative importance of viscous forces 
compared to surface tension can be quantified by a non-dimensional flow rate,
$Q = \mu^* U^*_0/\sigma^*$, where $\mu^*$ is the viscosity of the suspending
fluid, $U^*_0$ is the average flow velocity and $\sigma^*$ is the surface tension
at the interface between the suspending fluid and gas bubble.
In the absence of
fluid inertia, rigid spherical particles do not migrate normal to the direction
of flow\cite{CoxMasonReview1971} and hence, as highlighted by many
authors, the deformation of
droplets or bubbles is an essential part of their
underlying migration mechanism in slow flows\cite{LealReview1980,Stan_etal2011}.
For pressure-driven flow, in geometrically symmetric channels
a deformable object will migrate away from the walls
towards the centre of the channel\cite{KarnisMason1967}.
Hence, the development of passive sorting techniques that can take
advantage of the interaction between surface tension and viscous
forces requires a means to induce new geometrically distinct,
off-centre, propagation modes
where bubbles or droplets displace steadily near the channel walls 
and parallel to the fluid flow. 

One method for passively controlling
bubble or droplet location in the absence of
inertia is to induce geometric variation within the channel
cross-section. The introduction of grooves and holes into the top
surface of a microchannel allows droplets or vesicles to reduce their surface
energy by expanding into the available space\cite{Abbyad2011,Yamada2014}. Thus,
by placing these grooves off-centre, droplets are anchored to the groove
and can be guided to desired
locations under moderate flow rates\cite{Abbyad2011}. Furthermore,  multiple
grooves of different widths can be used to
sort droplets by their size or capillary
number\cite{Yoon_etal2014}. The
fundamental anchoring mechanism is based on modification of the
capillary-static solution, a mechanism that operates for
bubbles at both millimetric and micrometric
scales\cite{Dawson_etal2013}. If the bubble is forced instead to contract
via the introduction of a localised obstacle, capillary forces will
drive the bubble away from the constricted region. This effect has recently
been used to facilitate faster fluorescence-activated
sorting by Sciambi \& Abate\cite{Sciambi2015},
who used what they termed a ``gapped
divider'' that occluded approximately one third of the channel's
height. 

In this work,
we study the propagation of finite air bubbles using silicone oil as a carrier
fluid in a channel with a rectangular cross-section of high aspect
ratio
(width of the channel over its
height $\alpha\gg 1$) in which a small centred height constriction that we term
a ``rail'' is introduced in the longitudinal direction of the
channel. We restrict attention to bubbles with diameters (when viewed from above) between 70\% and
180\% of the width of the rail. Our selected channel geometry is sometimes known as a
Hele-Shaw cell, where  in the absence of height variation, only centred bubbles will propagate stably for non-zero values of the surface
tension\cite{TanveerSaffman1987}. In this geometry, static bubbles
that do not span the full channel width are neutrally stable and can
be located anywhere within the cross-section provided that they do not
interact with the sidewalls. Thus, there must be a 
dynamic (flow-induced) restoring mechanism which ensures
that off-centre bubbles return to the centreline of the
channel. The restoring mechanism involves an interplay between the
viscous pressure gradient, droplet shape and capillary pressure drop
over the interface and does not appear to have been comprehensively 
elucidated in the
Hele-Shaw-cell literature, although it is fundamentally the same as mechanisms
operating in two-dimensional\cite{Richardson1973} and
three-dimensional\cite{ChanLeal_etal1979} Stokes flows.
As described above, the introduction of a central rail leads to
modification of the capillary-static solutions such that 
bubbles are driven
away from the centre of the channel for low to moderate flow
rates\cite{Sciambi2015}. However,
higher flow rates offer the possibility of dynamic stabilisation
of the central
``on-rail'' propagation mode through a similar interaction between viscous
 and surface tension forces to that in the unoccluded Hele-Shaw
 cell. The stabilisation of the ``on-rail'' propagation
 mode at high flow rates is the principal focus of this paper.

\begin{figure}
\includegraphics[width=\columnwidth]{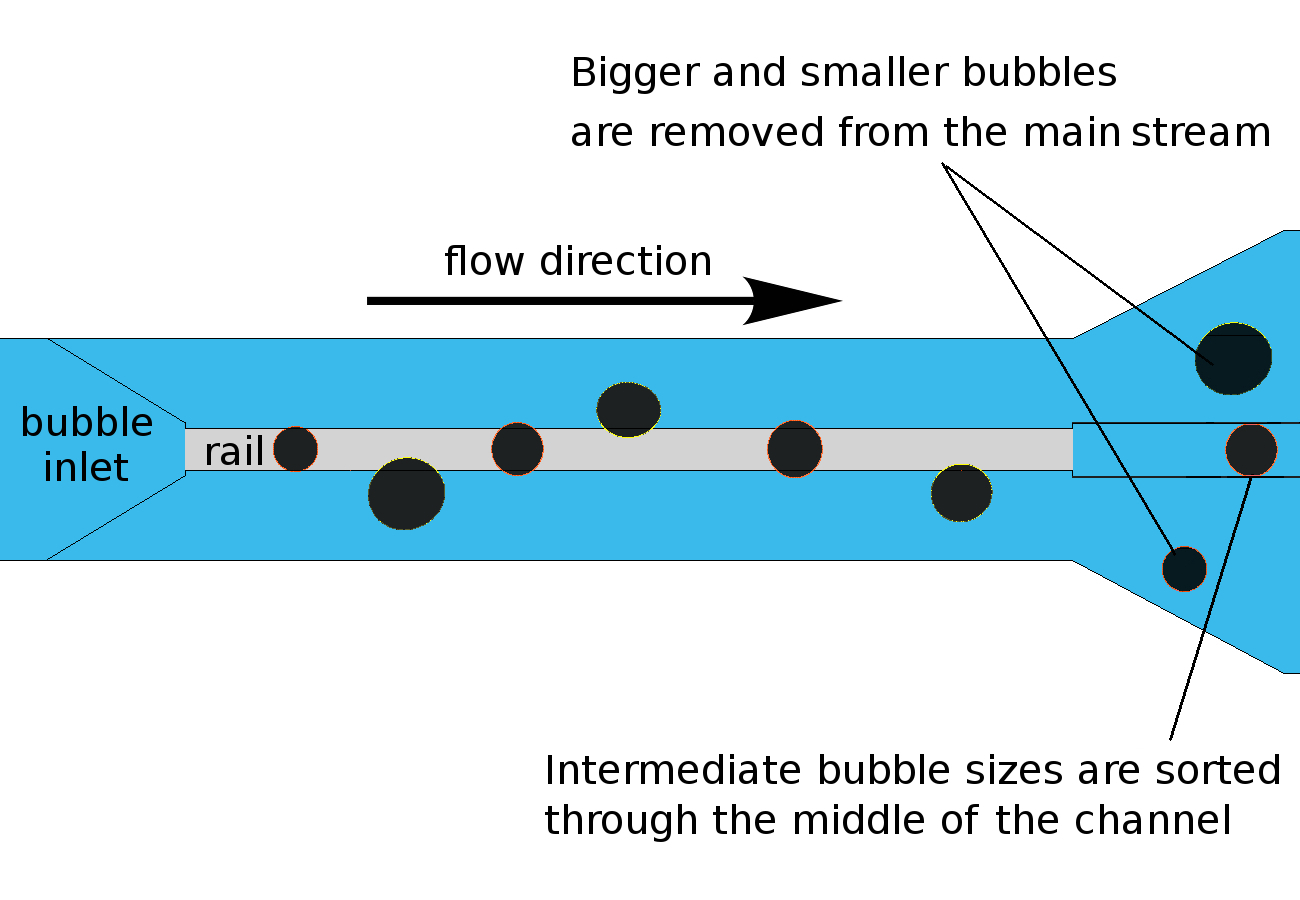}
\caption{In a channel of rectangular cross section bubbles typically
  propagate along the centreline. Upon the introduction of a centred rail 
  within such a channel, we find that near a critical flow rate 
  only a limited range of bubble sizes can propagate along
  the centreline, which provides a passive (geometric)
  mechanism for sorting bubbles by size. The critical flow rate and 
  range of selected bubble sizes are set by the rail width.
}
\label{sorting}
\end{figure}

In fact, our
experimental results reveal that stable propagation over the rail is
only possible for a limited range of bubble sizes. We find that the critical flow-rate for stable
propagation depends
non-monotonically on bubble size and that for a fixed flow rate marginally larger than the minimum critical flow rate, only a
narrow range of bubble sizes will propagate stably over the
rail.
Numerical computations using a depth-averaged model
reveal that this feature arises through the
evolving location of a symmetry-breaking bifurcation as a function of
bubble size and, moreover, that the dynamic stabilisation
of these intermediate-sized bubbles arises through a non-trivial interaction
between both viscous and surface-tension forces. In contrast, bubbles
of all sizes considered can always propagate stably in an off-centred position.
Thus, the
introduction of the rail can be used to design a passive bubble
sorting device that selects a restricted range of bubble sizes,
as sketched in figure \ref{sorting}.

In sections \ref{expmeth} and \ref{nummod} we describe our
experimental setup and numerical model, respectively. In section
\ref{sec:results}, we describe our results starting with the experimental evidence 
supporting passive bubble sorting. This is followed by a comparison with the results of our
depth-averaged model, which in turn provides an explanation of the mechanisms underlying the process. Finally, we draw our conclusions in section \ref{conc}.

\section{Experimental methods}\label{expmeth}

  \begin{figure}[h]
    \begin{center}
     \includegraphics[width=\columnwidth]{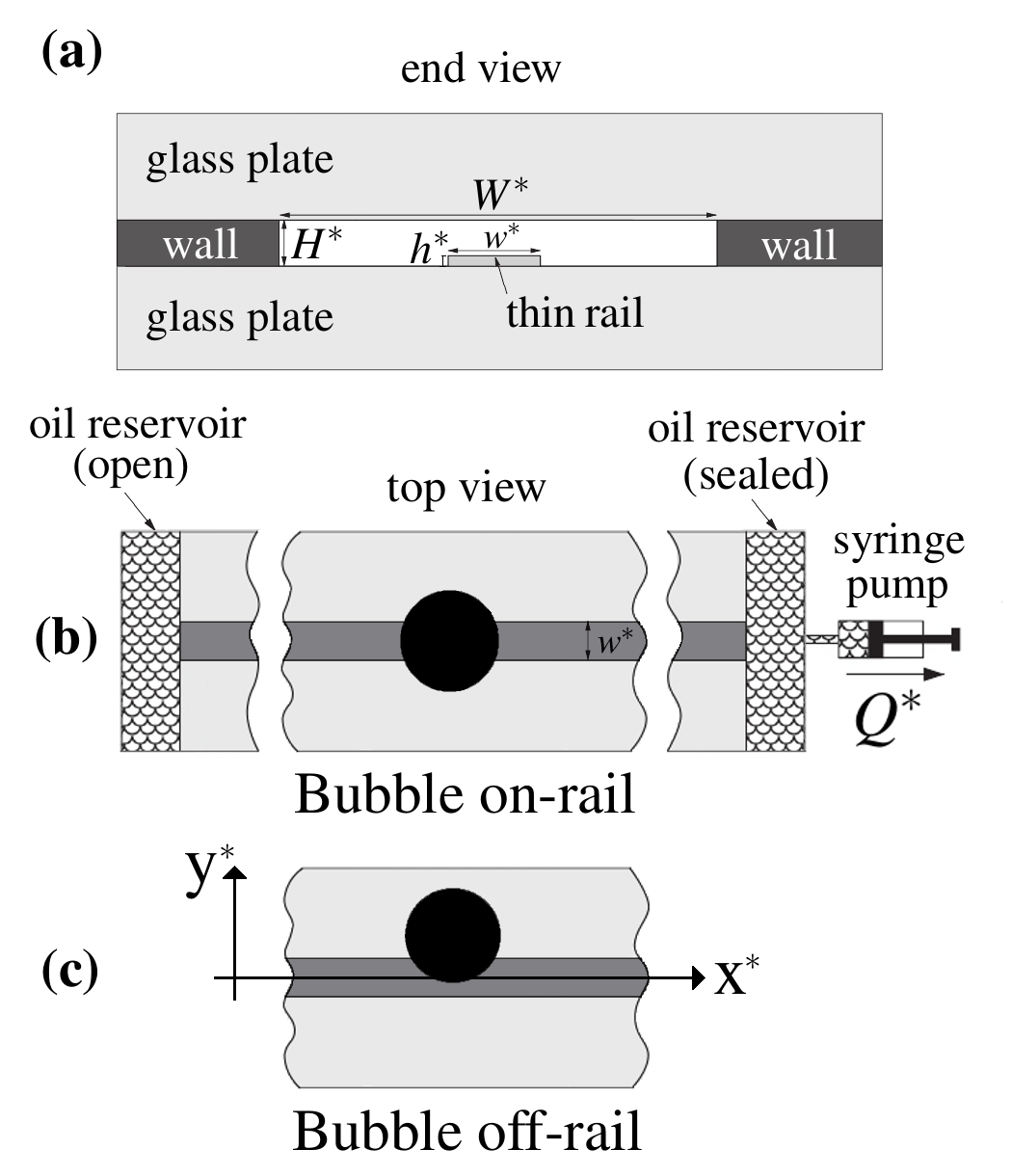}
    \end{center}
    \vspace{-0.5cm}
    \caption{\small  Schematic diagram of the experimental system. (a)
      Cross-sectional view of the channel with a prescribed depth
      variation: an axially uniform, thin rail is positioned on the
      bottom boundary, symmetrically about the centreline of the
      channel. (b) Top view of the experimental system. An air bubble
      is propagated in an on-rail position by withdrawing liquid with
      constant volume-flux from one end of the channel using a syringe
      pump. (c) Schematic diagram of a bubble propagating in an
      off-rail position, indicating the Cartesian coordinate system
      used.}\label{fig:Exp_setup}
  \end{figure}

A schematic diagram of the experimental setup is shown in figure
\ref{fig:Exp_setup}. Note that all dimensional variables are starred. 
The flow channel was made of two parallel glass
plates of dimensions $60$ cm $\times$ $10$ cm $\times$ $2$ cm
separated by brass sheets of uniform height $H^*=1$ mm, so that the
depth of the channel was accurate to within $0.1\%$. The width of the
channel was set to $W^*=30\pm 0.1$ mm, and thus the aspect ratio was
$\alpha = W^*/H^*=30$. A prescribed depth profile was introduced by
bonding a rectangular strip of polypropylene film of thickness
$h^*=24\pm1$ $\mu$m (i.e. $2.4$ \% of the total height of the channel)
along the centreline of the bottom boundary, which hereinafter will be
referred to as the rail (figure \ref{fig:Exp_setup}a). Two different
rail widths were used:  $w^*=6.9\pm0.1$ mm and $10.7\pm0.1$ mm. The
positional accuracy of the rail was such that the widths of the
full-height channels on both sides of the rail differed by less than
$1$ \% ($<100$ $\mu$m) over the entire length of the channel. The
channel was levelled horizontally to within $\pm0.05^{\circ}$.

A syringe pump (KDS210) was used to fill the channel with 
silicone oil  (Basildon Chemicals Ltd.) of
viscosity $\mu^*=5\times10^{-2}$ Pa s, density $\rho^*=961$ kg m$^{-3}$
and surface tension $\sigma^*=2.1\times10^{-2}$ N m$^{-1}$ at the
laboratory temperature of $21^\circ$C, via the
sealed reservoir at one end of the channel (figure \ref{fig:Exp_setup}b). At the other end of the
channel, oil was allowed to accumulate in an open reservoir. A single
bubble was introduced in the channel with a $5$~ml glass syringe
(Hamilton Gastight) connected to a $10$~cm long needle with a diameter
of $0.8$~mm, inserted into the open end of the channel. The
bubble was then propagated by withdrawing oil using the syringe pump
at a constant value of the flow rate $Q^*$ in the range $1\leq
Q^*\leq 50$ ml/min. Once the bubble had travelled the length of the
channel, it was returned to the inlet by infusing oil back into the
channel. At the start of each experiment, the bubble was approximately
at rest and positioned either centrally in the channel (on-rail,
figure \ref{fig:Exp_setup}b), or asymmetrically about the centreline
of the channel (off-rail, figure \ref{fig:Exp_setup}c). For the range
of  bubble sizes investigated, the on-rail initial position was
unstable, and thus it could only be imposed transiently (see
Supplementary Material for details).

The propagation of bubbles was recorded with a DALSA Genie TS-M3500
camera with a $35$ mm {\it f}/1.4 lens (Carl Zeiss T$^{*}$ Distagon)
mounted at a distance of $0.94$ m above the experiment. The channel
was back-lit with white light of uniform intensity provided by a
custom-made  LED light box aligned under the channel. The field of
view of the camera was a channel section of dimensions $262.0$ mm
$\times$ $30.0$ mm, and the image resolution of $1920\times 220$
pixels yielded $136.4$ $\mu$m/px. Depending on the value of $Q^*$,
time-sequences of $400\geq N\geq 120$ images were recorded at frame
rates $1\leq f^*\leq10$ frames per second. An image processing algorithm
developed in Matlab (MATLAB R2014a) was used to extract the outline of
the propagating bubbles \cite{FrancoGomez_etal2016} as a function of time. The edge of the
bubble coincides with the external edge of the outline, 
determined to within $0.14$~mm. The
projected area $A^*$ of the bubble was calculated from the
extracted outline. The centroid of this projected area was then determined to yield its displacement from the centreline of the channel, $y_c^*$. The size of the bubble was characterised by its
diameter $D^*=2 \sqrt{A^*/\pi}$ at rest off the rail. The range of
diameters investigated was between $18\%$ and $49\%$ of the channel
width. Given that the depth of the channel was $1/30$ of its width, this meant that the bubbles were strongly confined into a
quasi-two-dimensional shape with thin liquid films separating the
bubble interface from the top and bottom boundaries of the
channel. Hereafter, we will quote diameters relative to the width of
the rail, $D=D^*/w^*$. 
The evolution of
the dimensionless centroid position $y_c = 2 y_c^*/W^*$ in time was
examined in order to determine which propagating bubbles reached a
steady state within the length of the channel.  
Results will be discussed in terms of a non-dimensional flow
rate $Q=\mu^*  U^{*}_{\rm 0}/\sigma^*$,  where $U^{*}_{\rm 0}=Q^* / (W^*H^*)$ 
is the average speed of the fluid flow within
the channel in the absence of the rail. The
Reynolds number of the channel flow remained within the range $0.16
\leq Re_\infty = (\rho^* U^{*}_{\rm 0)} d_h^*)/\mu^* \leq 0.39$, where
$d_h^* = (2 W^* H^*) / (W^*+ H^*) = 1.94$~mm is the hydraulic diameter of
the channel, suggesting that inertial forces were negligible in the
experiments \cite{Amini_etal2014}.

\section{Depth-averaged model}
\label{nummod}

 \begin{figure}[h]
    \begin{center}
     \includegraphics[width=\columnwidth]{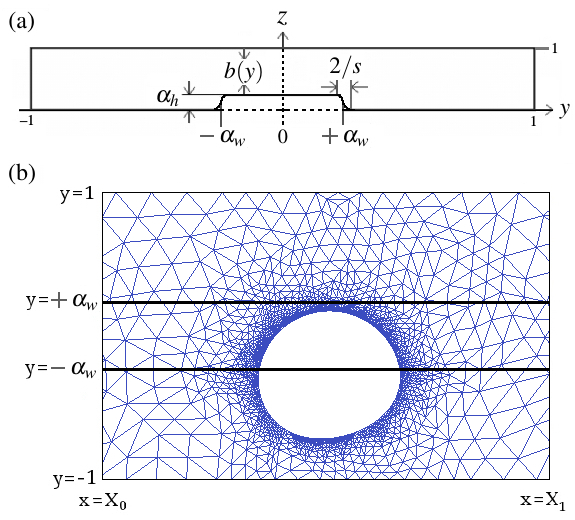}
    \end{center}
    \vspace{-0.5cm}
    \caption{\small (a) Cross-section
      of the channel including a rail with relative width $\alpha_w$,
      relative height $\alpha_{h}$ and sharpness $s$. The channel
      depth profile is given by $b(y)$. (b) Top view of the channel,
      showing a typical finite element mesh and bubble contour.
      Solid black lines at $\alpha_w$ and
      $-\alpha_w$ represent the edges of the rail. Both (a) and
      (b) are plotted in a dimensionless coordinate system.}\label{fig:BubbleMeshRailProf}
  \end{figure}
  
We model the motion of finite bubbles on a rail by extending our
two-dimensional lubrication model\cite{Thompson2014}, initially developed
to describe propagation of air fingers (open-ended bubbles) in similar
channels. The model described below
was later shown to be quantitatively accurate in
channels of sufficiently large aspect ratio $\alpha \ge
40$, for thin rails with $\alpha_h<0.12$, and moderate values of the
capillary number $Ca<0.012$\cite{FrancoGomez_etal2016}. 

The geometry of a typical 
channel cross-section is shown in figure
\ref{fig:BubbleMeshRailProf}a. We introduce a Cartesian coordinate 
system aligned with the channel such that $x$ is the axial coordinate and $y$
and $z$ span the cross-section. 
The $x$ and $y$ (width) coordinates have been non-dimensionalised by
$W^*/2$, whereas the $z$ coordinate (height) has been non-dimensionalised by
$H^*$. The rail is represented by a smooth
tanh-profile leading to a channel depth profile of the form
  \begin{equation}\label{DepthProfile}
     b(y) = 1 - \frac{\alpha_h}{2}[\tanh s(y+\alpha_w)-\tanh s(y-\alpha_w)],
  \end{equation}
where $s$ is the sharpness of profile edges, $\alpha_h=h^*/H^*$ is the fractional height of the obstacle, and $\alpha_w$ is the fractional width of the rail at half its maximum height.

We define a reference velocity scale as $U_0^* = Q^*/(W^* H^*)$, and non-dimensionalise the depth-averaged horizontal velocity $\bf{u^*}$  on the scale $U_0^*$, pressure $p^*$ on the scale $6\alpha \mu^*  U_0^*/(H_0^*)^2$ and time on the scale $W^*/(2 U_0^*)$.

 After applying the lubrication approximation
\cite{Reynolds1886}, the governing equation for the viscous,
incompressible fluid in the
frame of reference moving with the bubble tip $\mathbf{U}_b = (U_b,0)$,
where $U_{b} = U_{b}^{*}/U_{0}^{*}$, is
 \begin{equation}\label{LaplaceEq}
    \bm{u} = -b^2 \bm{\nabla} p, \quad \bm{\nabla}\cdot(b(y)^3 \bm{\nabla} p)=0 \quad \mathrm{in}\quad  \Omega,
  \end{equation}
 where $\Omega$ denotes the fluid domain. The fluid domain is $X_0 \leq x \leq X_1$, $-1 \leq y \leq 1$, excluding 
the region occupied by the bubble, where $X_0$ and $X_1$ are truncation 
coordinates behind and ahead of the centre of the bubble (figure \ref{fig:BubbleMeshRailProf}b).

The conditions at the bubble interface $\mathbf{R}=(x,y)$ and on the
channel boundaries are 
  \begin{equation}\label{PressureJump}
       p_{b}-p=\frac{1}{3\alpha 
         Q}\left(\frac{1}{b(y)}+\frac{\kappa}{\alpha}\right)\quad
       \mathrm{on}\quad  \partial \Omega_b,
  \end{equation}
  \begin{equation}\label{VeloCont}
       {\bf\hat n}\cdot \frac{\partial \mathbf{R}}{\partial t}+{\bf\hat
         n}\cdot \mathbf{U_b}+b^2{\bf\hat n}\cdot \bm{\nabla} p=0\quad
       \mathrm{on}\quad  \partial \Omega_b,
  \end{equation} 
  \begin{align}\label{PressWallEnds}
       \frac{\partial p}{\partial y}=0\quad \mathrm{on}\quad y=\pm 1.
  \end{align} 
  where $\partial \Omega_b$ denotes the bubble boundary with unit
  normal ${\bf \hat n}$ directed into the fluid. 

  The dynamic boundary condition (\ref{PressureJump}) is the
  non-dimensional form of the Young-Laplace equation, where $p_{b}$
  is the pressure inside the bubble, $Q=\mu^* U_0^*/
  \sigma^*$ is the dimensionless flow rate based on the net 
  volume flux, and $\kappa$ is the dimensionless
  curvature of the interface in the $(x,y)$ plane. We
  assume for both kinematic and dynamic boundary conditions that the bubble occupies the full height of the channel,
  which neglects the effects of the thin films known to develop above
  and below the bubble.

  The
  interfacial conditions in the moving frame have been previously
  derived\cite{FrancoGomez_etal2016} and the only
  difference from the interfacial conditions in the fixed
  lab frame\cite{Thompson2014}, is in equation
  (\ref{VeloCont}), which adds the velocity of the
  frame moving with the tip of the bubble, $\mathbf{u}=-\mathbf{U}_b-b^2 
\bm{\nabla}
  p$, into the kinematic boundary condition
  $\displaystyle{\frac{\partial \mathbf{R}}{\partial t}\cdot 
\mathbf{\hat{n}}=\mathbf{u}\cdot\mathbf{\hat{n}}}$.

  In the experiments, the control parameter is the volume flux $Q^*$ imposed by the syringe pump. In the model, we set the fluid pressure to zero at the rear of the domain, and applying a pressure gradient $-\Lambda\bf{e}_x$ far ahead of the bubble tip, where the value of $\Lambda$ is chosen to satisfy the dimensionless volume constraint:
\begin{equation}
 \frac{1}{2}\int_{-1}^1 b(y) \mathbf{u}\cdot \mathbf{e}_x \, \mathrm{d}y = 1.
\end{equation}
By recalling that $\mathbf{u}\cdot \mathbf{e}_x= - b^2 p_x\rightarrow -b^2 \Lambda$,
 we can in fact solve for $\Lambda$ in terms of an integral:
\begin{equation}
\Lambda = \frac{2}{\int_{-1}^1 b^3(y)\,\mathrm{d}y}.
\end{equation}
The dimensionless constant $\Lambda$ is a function of channel geometry only, taking the value $1$ for a rectangular channel, and increasing with obstacle height.

The dynamic boundary condition (\ref{PressureJump}) does not
  include the correction
  factor of $\pi/4$ first derived by \citet{ParkHomsy1984} required to
  match to the static solution in large aspect-ratio channels in the
  absence of any depth variations. A constant factor of this form can be accommodated
  within our model by rescaling $\alpha$ and $Q$ as discussed by
  \citet{FrancoGomez_etal2016}, and so would not qualitatively alter the results; however \citeauthor{FrancoGomez_etal2016} found that its inclusion actually increased the error between model results and the experimental data for air fingers at the typical flowrates considered here. A bigger
  difficulty is that the correction factor
  itself could be altered by the presence of the rail and the flow rate, and can display significant spatial variation even without topographic variations \cite{ZhuGallaire2016}.
  In principle, we could dynamically adjust the curvature term according to
  measured film thicknesses, but this is not a predictive model and is complicated by 
  the existence of multiple solutions for bubble shape and hence for film thickness.
  Any such adjustment would therefore be
  somewhat \textsl{ad hoc} and we prefer to
  keep the model as simple as possible in
  order to explore underlying mechanisms, rather than pursuing exact
  quantitative agreement. Nevertheless, we should remark
  that \citet{deLozar2008} did find quantitative agreement between the
  uncorrected model presented here and three-dimensional 
  Stokes flow simulations of air fingers
  for channel aspect ratios greater than or equal to $8$ and capillary
  numbers above $0.01$, for channels without obstacles.

  A second adjustment to the basic depth-averaged model that has been proposed\cite{Nagel2014} is the use of Brinkman equations, involving the ad-hoc inclusion of higher order lateral diffusion terms. The advantage of the Brinkman equations is that both the no-slip boundary condition on the wall and the tangential stress boundary condition on the bubble boundary can be satisfied. The bubble diameters considered here are much smaller than the channel width, so wall slip is unlikely to be significant. However, it is unclear how significant the tangential stress boundary layer would be, nor is it clear how the Brinkman equations should be modified by the presence of the obstacle. Again, we will omit Brinkman terms here for the sake of simplicity; the question of the validity of various depth-averaged models, the importance of dynamic film thickness, and how these interact with topographic variations, would best be explored by detailed full three-dimensional numerical computations.


    The term $\partial \mathbf{R}/\partial t$ is the only time derivative in the problem and
  drives the unsteady evolution of the bubble. In this work we focus on
  steady solutions computed by setting $\partial \mathbf{R}/\partial
  t=0$, but we also present results using time-dependent simulations
  in figures \ref{fig:transient_evolution}, \ref{fig:PertCentNumStepsEffects} and
  \ref{fig:PressJumpCentallBubbs}.
  Note that the time-derivative term also features in the stability
  analysis of the steady states.

  The bubble velocity $U_b$ is an unknown with the associated 
  constraint that the centroid of the bubble is fixed at $x=0$.
  The fluid pressure is fixed to zero far ahead of the bubble, which
  means that the bubble pressure $p_{b}$ is also an unknown with the
  associated constraint that the bubble volume must remain a
  prescribed constant, $V_{0}$, during the evolution 
  \begin{equation}
   \iint_{\Omega_{b}} b(y) \mbox{d}A = V_{0}, \label{VolCons}
  \end{equation} 
  assuming that the bubble occupies the full height of the channel.
Note that the numerical values for $y_c$ are calculated using the 
height-weighted centroid position (see Supplementary Material); this 
differs from the experimental definition by at most 1.1\%.

  The model is solved using the finite element library
 {\it oomph-lib}\cite{HeilHazel2006} and implementation details are
 given in the supplementary material as well as in our previous papers
 \cite{Thompson2014,FrancoGomez_etal2016}.

\section{Results}\label{sec:results}

\begin{figure*}
    \begin{center}
       \includegraphics[width=\textwidth]{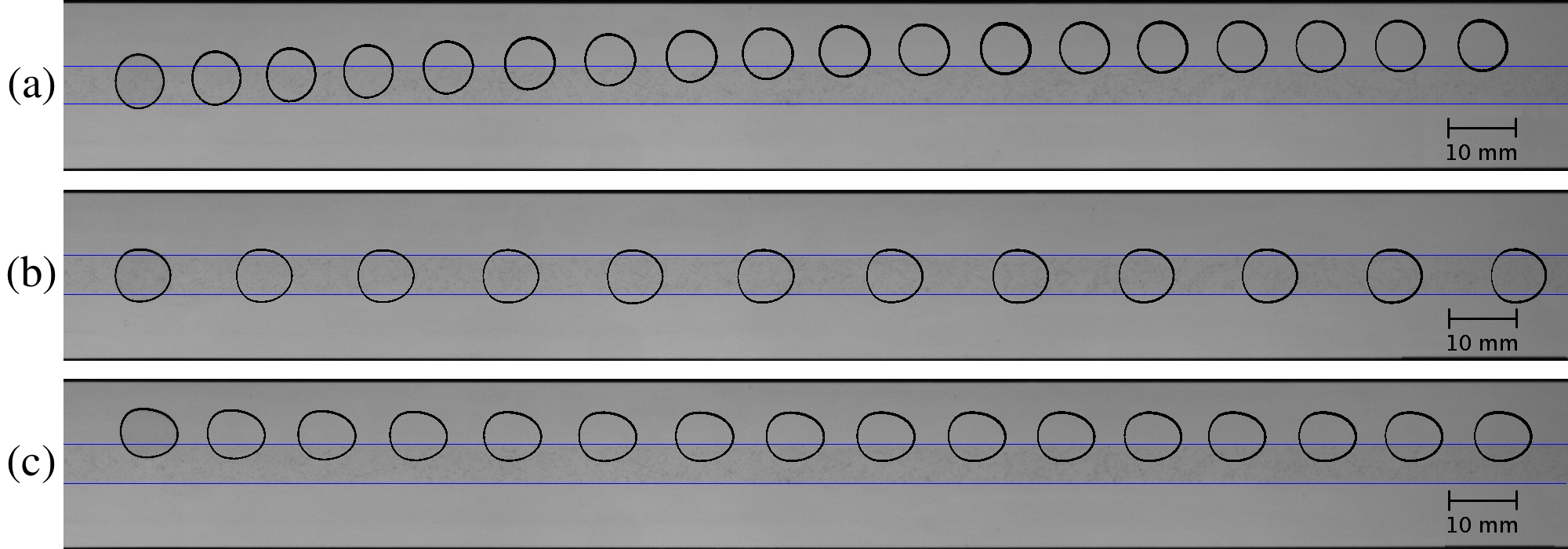}
    \end{center}
    \caption{\small Time-lag images of a propagating bubble with
      diameter $D=D^*/w^*=1.33$  in a channel with aspect ratio
      $\alpha=30$, rail width $w^*=6.9\pm0.1$ mm and rail height
      $h^*=24.0$ $\mu$m. The solid blue lines delineate the edges of
      the rail. (a)  $Q^*=3.0$ ml/min: the bubble starts in the
      on-rail position and is displaced towards an off-rail stable
      state (images taken every $5$ s) . (b) $Q^*=19.0$ ml/min: the
      bubble starts in the on-rail position and propagates on-rail
      (images taken every $1$ s). (c) $Q^*=13.0$ ml/min: the bubble
      starts in the off-rail position and propagates off-rail
      (snapshots every $1$ s).}\label{fig:Cen_flowrateSchem}
  \end{figure*} 
  
\subsection{On and off-rail bubble propagation}

As described in the introduction, in a channel of uniform depth bubbles are 
transported along the centreline ($y=0$) of the channel
\cite{GoldsmithMason1962,Stan_etal2011}
and adopt states that are symmetric about the centreline.
In contrast, we have found that the introduction of a thin
 rail along the bottom of
the channel enables three modes of propagation: an on-rail symmetric
mode analogous to bubble propagation in channels of uniform
depth; and two asymmetric off-rail modes, one on each side
of the rail. In a given experimental system, only one of the
off-rail modes is generally observed, as illustrated in figure
\ref{fig:Cen_flowrateSchem}, because of a small, unavoidable bias in
the experimental channel. 

\begin{figure}
 \includegraphics[width=\columnwidth]{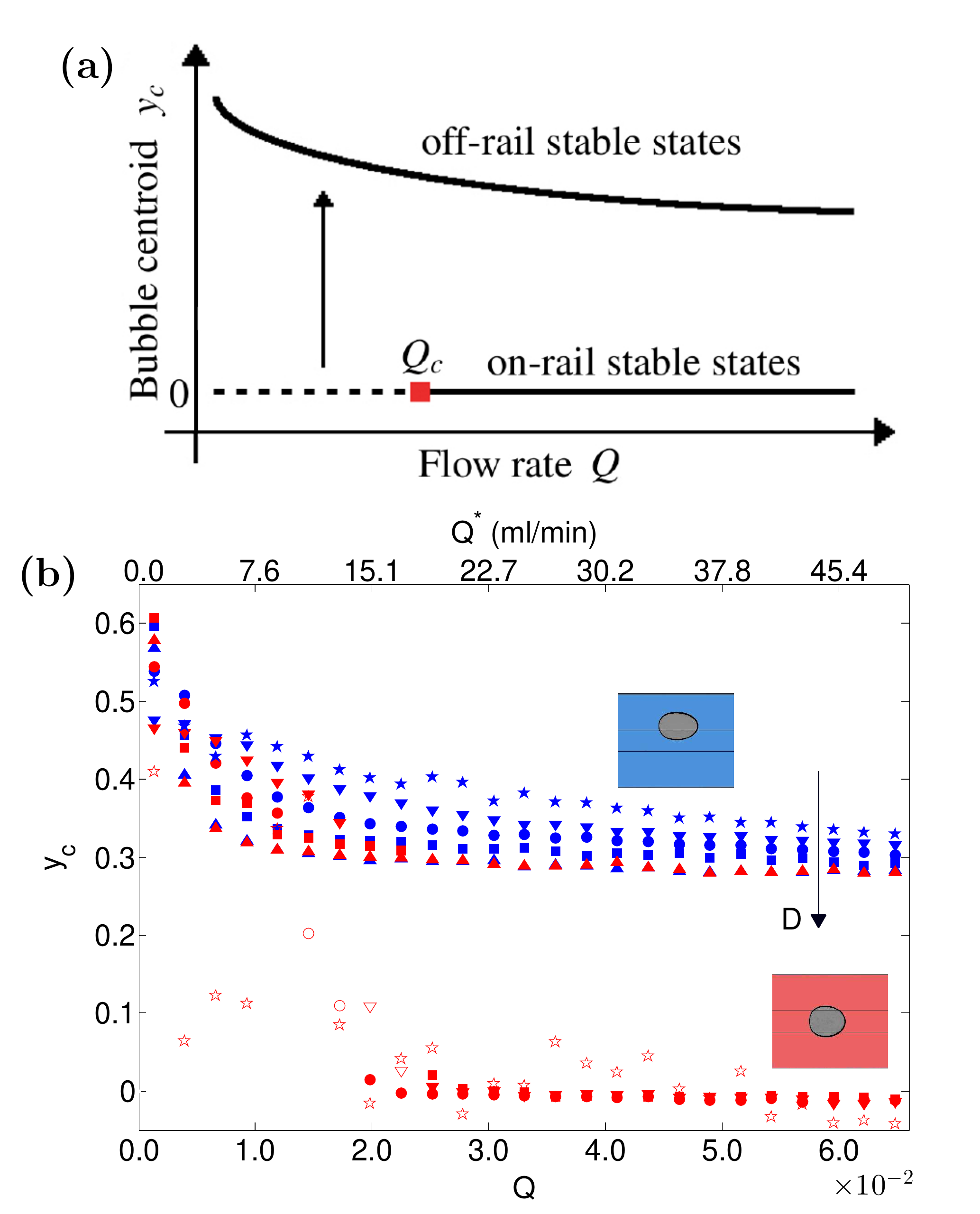}
  \caption{\small (a) Schematic diagram summarising the experimental
  observations illustrated in figure \ref{fig:Cen_flowrateSchem}. (b)
  Quantitative data corresponding to the schematic diagram: the 
  centroid position $y_c=2y^*_c/W^*$ of steadily propagating bubbles of
  different sizes as a function of flow rate $Q^*$ in the experimental
  channel with a rail of width $w^*=6.9\pm0.1$~mm and height
  $h^*=24.0$ $\mu$m. The bubble size is quantified by the diameter of
  the bubble at rest, relative to the width of the rail, $D=D^{*}/w^{*}$,
   in the on-rail position: ({\bf$\bigstar$})
  $D=0.77$, (\textcolor{black}{\bf$\blacktriangledown$}) $D=1.04$,
  (\textcolor{black}{\Large\bf$\bullet$}) $D=1.33$,
  (\textcolor{black}{$\blacksquare$}) $D=1.65$ and ({\bf
    $\blacktriangle$}) $D=1.81$. Red (blue) symbols correspond to
   initial conditions being on-rail (off-rail).
  Open symbols denote experiments where
  a steadily propagating state was not reached within the length of
  the channel. The cluster of open stars in the vicinity of $y_c=0$
  denote experiments on the smallest bubble with $D=0.77$, for which a
  steady on-rail state was not observed, but where the migration
  towards the off-rail state was very slow.}\label{fig:ExpBistability}
  \end{figure}

The stability of the different propagation modes depends on the
channel geometry, bubble size and flow rate, but the mode selected 
also depends on the initial conditions. If a bubble of diameter
$D = D^{*}/w^{*} = 1.33$  is started from an on-rail
position, it migrates towards an off-rail position
for low flow rates (figure \ref{fig:Cen_flowrateSchem}a), whereas
above a critical flow rate $Q_c$, the bubble propagates steadily on
the rail (figure \ref{fig:Cen_flowrateSchem}b). By contrast, if
started in an off-rail position, the bubble is transported steadily in
an off-rail position for all values of $Q$ investigated (figure
\ref{fig:Cen_flowrateSchem}c). Hence, for $Q > Q_c$, the device
exhibits bistability, with on-rail (off-rail) propagation reached from
on-rail (off-rail)  initial conditions, respectively, while for
$Q<Q_c$, the on-rail mode of bubble propagation is unstable.  This
scenario is summarised in a schematic bifurcation diagram in figure
\ref{fig:ExpBistability}a. 

 Quantification of the existence of the on and off-rail modes of propagation is 
provided in figure \ref{fig:ExpBistability}b, where
the centroid position $y_c$ of steadily propagating bubbles is plotted
as a function of flow rate for a range of static bubble diameters
relative to the rail width, $0.77< D=D^*/w^* <1.81$, for the case of rail
width $w^{*}=6.9$~mm. Red (blue)
symbols correspond to on-rail (off-rail) initial conditions,
respectively. For off-rail initial conditions all bubbles propagated in
the steady off-rail state. For on-rail initial conditions, the
smallest bubble $D=0.77$ does not attain a steadily propagating
state. Bubbles with
$1.04 \leq D \leq 1.65$ propagate steadily on the rail for
sufficiently large flow rates. The largest bubble $D=1.81$ did 
not propagate steadily on the rail, but instead rapidly migrated sideways
to attain a steady off-rail propagation state (figure
\ref{fig:Cen_flowrateSchem}a). 
Open symbols indicate bubbles that have not reached a steady state
within the visualisation window. In the main, these are bubbles near
the stability threshold, with the exception of the
smallest bubble, $D=0.77$,  for which the migration to an off-rail
state was very slow.   

\subsection{On-rail stability tongues: a concept for sorting bubbles by size}
\label{exptongue}

   \begin{figure}[t]
    \begin{center} 
       \includegraphics[width=\columnwidth]{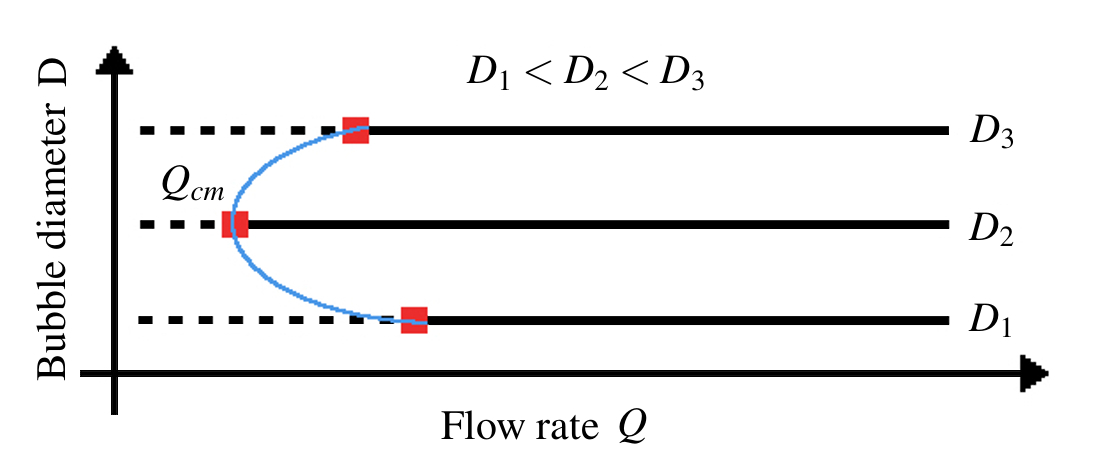}
     \end{center}
     \vspace{-0.5cm}
     \caption{\small Schematic diagram summarising the stability of
       the on-rail mode of propagation as a function of flow rate for
       three different bubble diameters. The red squares indicate the
       critical flow $Q_c$ rate beyond which the on-rail mode is
       stable for each bubble. The variation of $Q_c$ with bubble
       diameter observed  in figure \ref{fig:ExpBistability} is
       non-monotonic, resulting in a tongue-shaped stability boundary
       between unstable and stable on-rail modes in the parameter
       plane spanned by bubble diameter and flow rate. The minimum
       value of the critical flow rate $Q_{cm}$ occurs for an
       intermediate bubble size.}\label{fig:ExpBistability2}
  \end{figure}  

Figure \ref{fig:ExpBistability}b suggests that $Q_c$, the smallest
value of flow rate for which an initially centred bubble remains on
the rail during propagation to the end of the channel, varies
non-monotonically with bubble diameter. Inspection of the data
suggests a tongue-shaped stability boundary in the parameter plane spanning
bubble diameter and flow rate, with a minimum value of the critical
flow rate, $Q_{cm}$, for an intermediate bubble size, illustrated
schematically in figure \ref{fig:ExpBistability2}. Hence, for a flow
rate marginally larger than $Q_{cm}$, only those bubbles within a narrow
band of sizes can propagate steadily on-rail; bubbles of other
sizes will migrate to off-rail positions. 

\begin{figure*}[t]
\includegraphics[width=\textwidth]{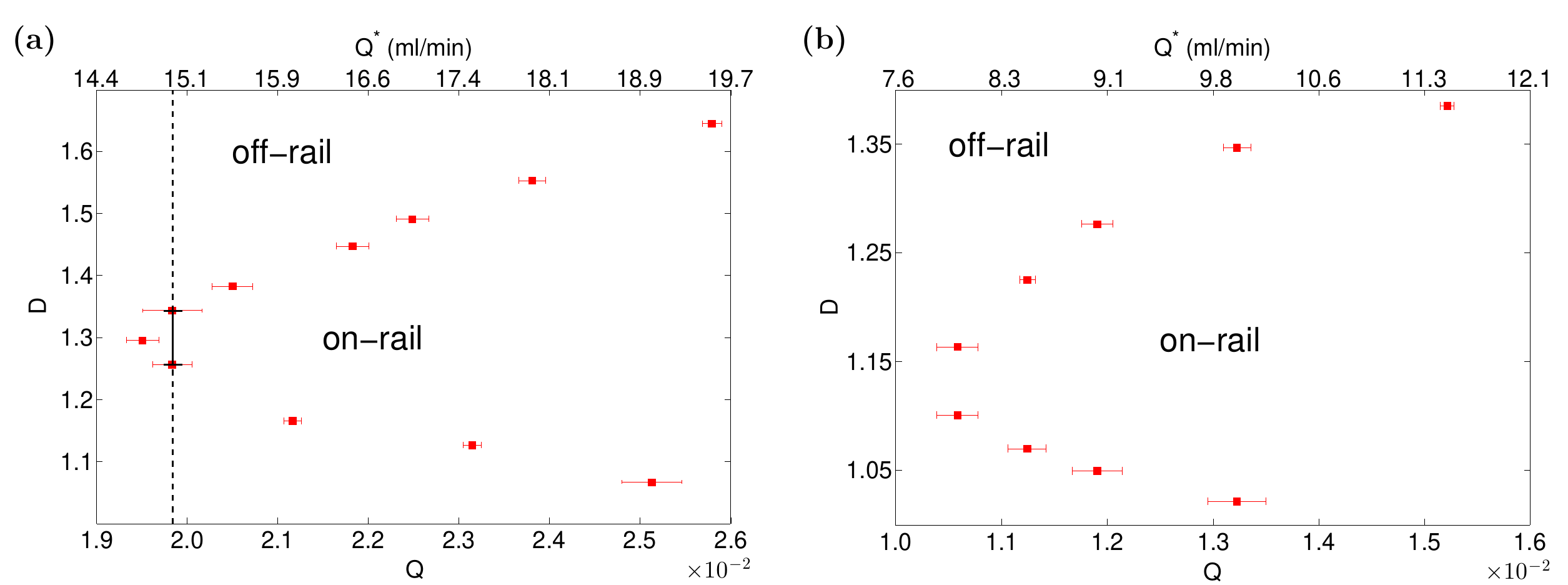}
\caption{\small Experimental stability boundaries in the parameter plane spanning
  bubble area and flow rate, which demarcate regions of bistability
  (inside the tongue) from regions of off-rail propagation (outside
  the tongue). (a) $w^*=6.9 \pm 0.1 $~mm, where $Q_{cm}^*=
  14.7$~ml/min for $D^*=8.94$~mm. The dashed line highlights the narrowest band of bubble diameters that may be stabilised robustly in this device. (b) $w^*=10.7\pm 0.1 $~mm, where
  $Q_{cm}^*= 8.0$~ml/min for
  $D^*=12.12$~mm.}\label{fig:CentreAreaBiTime}
  \end{figure*}

 The experimental stability boundaries, showing $Q_c$ as a function of
 bubble diameter, are presented  in figure
 \ref{fig:CentreAreaBiTime}. The range of bubble diameters tested for
 $w^*=6.9$~mm (figure \ref{fig:CentreAreaBiTime}a) and $w^*=10.7$~mm
 (figure \ref{fig:CentreAreaBiTime}b) was $1.07 \le D \le 1.65$ and
 $1.02 \le D \le 1.39$, respectively. The stability
 boundaries demarcate regions of stable on-rail bubble propagation,
 which lie inside the tongues, from regions of stable off-rail
 propagation for bubbles set into motion in an on-rail position. Both
 graphs are qualitatively similar but exhibit different dimensional
 values of $Q_{cm}^*= 14.7$~ml/min and $Q_{cm}^* = 8.0$~ml/min
 associated with distinct (dimensional) bubble diameters $D=1.29$
 ($D^*=8.94$~mm) and $D=1.13$ ($D^*=12.12$~mm), respectively. The increase in the
 dimensional size of the bubble at $Q_{cm}$ with the widening of the
 rail suggests that the band of bubble sizes that can be segregated
 through propagation on the rail can be tuned by varying the rail
 width. 

 For the smaller bubble diameters, the time
required to reach a steady off-rail propagation mode from an on-rail
initial condition, was generally found to increase as the bubble size
decreased. For bubbles with $D=0.77$, a precise value of $Q_c^*$
could not be determined due to the length of the transient migration
of the bubbles towards a steady state, but a study of the time
evolution of the centroid position for increasing values of $Q^*$
suggests a value $43.0<Q_c^*<50$~ml/min.  These long durations and thus extended lengths of
migration could prevent the sorting of bubbles near the size thresholds
in practical devices. In all cases, however, the
duration of transient migration near the threshold $Q^*_c$ could be
reduced by biasing the symmetric on-rail initial condition, thus enabling segregation by size in channels of finite
length. For example, when the initial centroid displacement from the channel centreline was increased from the experimental tolerance of $500\; \mu$m to $1000\; \mu$m, the duration, and thus length, of transients were reduced by up to 40\%. 

Bubbles with larger diameters than those shown in figure
\ref{fig:CentreAreaBiTime}, i.e. $D > 1.65$ for $w^*=6.9$~mm, and $D
> 1.39$ for  $w^*=10.7$~mm, did not propagate steadily on-rail.
However, for bubbles large enough to span the entire width of the
channel in the static configuration, $D\alpha_{w} \approx 1$, a symmetric (on-rail) propagation 
mode can be supported at low values of the flow rate; above a critical 
flow rate, this symmetric mode loses stability to a pair of asymmetric propagation
states through either a supercritical or subcritical pitchfork
bifurcation depending on the rail height \cite{FrancoGomez_etal2016}. These observations highlight the
complex dependence of the
modes of propagation on the size of the bubble and hint at the rich
nonlinear dynamics of bubble propagation exhibited by this system.

\subsection{Comparison between experimental and numerical stability tongues}

  \begin{figure*}
     \includegraphics[width=\textwidth]{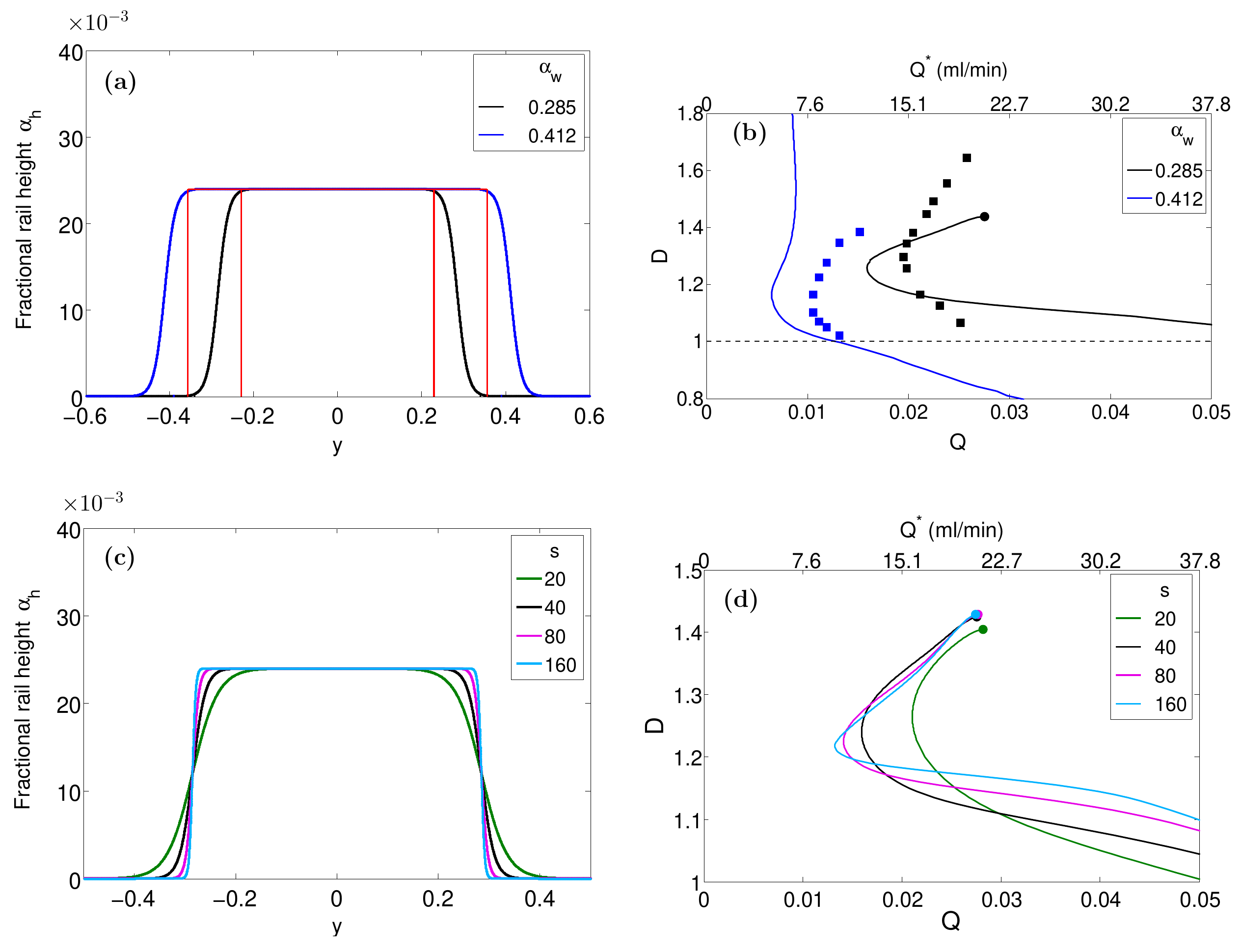}
     \caption{\small (a) Comparison between the model rail
       profiles with $\alpha_h=0.024$, shown with thick solid
       lines ($\alpha_w=0.285$, $s=40$ (black),
       $\alpha_w=0.310$, $s=40$ (blue)), and the corresponding experimental
       rail profiles measured by profilometry (red lines).
       (b) Comparison  for the rail geometries shown in (a) between the path of the pitchfork bifurcation that stabilises the on-rail mode of propagation (lines), plotted in the parameter plane spanned by flow rate and bubble diameter, and the experimental stability boundaries (symbols), previously plotted in figure \ref{fig:CentreAreaBiTime}.
       The horizontal dashed line marks the width of the rail, while the solid circles mark the ends of the pitchfork bifurcation paths, which are discussed in \ref{sec:biftrack}.
      (c)  Model rail profiles with varying levels of sharpness with $\alpha_w=0.285$. (d) Paths of the pitchfork bifurcation
       corresponding to the rail profiles in (c), showing the
       sensitivity of the results to the sharpness of the rail.  \label{fig:BifTrack}}
\end{figure*}

  We first 
evaluate the predictive capability of our depth-averaged model for finite 
bubbles with diameters
ranging between one and two rail widths. The model has
been shown to be quantitatively accurate when modelling
air finger propagation in
channels of larger aspect ratios, $\alpha \geq 40$, at relatively low
flow rates, so that liquid films above and below the finger may be 
neglected\cite{FrancoGomez_etal2016}. In these simulations, the
results were unaltered\cite{Thompson2014} for sufficiently sharp
changes in channel depth, $s \geq 40$ in equation (\ref{DepthProfile}). 
In the present study, however, we cannot rely on the quantitative
accuracy of the model because
the steady on-rail state is stabilised at relatively large values of
$Q$; the small finite bubbles used here may be more sensitive to details of the
rail geometry than the long air fingers; and the aspect ratio is only  $\alpha=30$. For large bubbles (with diameters approaching the channel width) at low flow rates ($Q\lesssim 0.01$), we have recently demonstrated quantitative agreement between the model and experimental data 
\cite{FrancoGomez_etal17b}

The rail height in the model was chosen based on micrometer
measurements of the height of the experimental rail, so that $\alpha_h=0.024$. Measurements of the
the height profile at six different
locations along the rail length using a step profilometer (DekTak
IIA), with a resolution of $0.1 \mu$m (see Supplementary Material), are consistent with the micrometer measurements, but they indicate that the top surface of the rail is rough with a
standard deviation of the height from its mean value of approximately
$3 \mu$m.  Moreover, the profilometry measurements show that the sidewalls of the rails are
approximately vertical. Hence, although we follow Franco-G\'omez et al. \cite{FrancoGomez_etal2016} in choosing a rail sharpness of $s=40$, and choosing $\alpha_w$ in order to match the top width of the rail, we also present a sensitivity analysis of the stability of the on-rail mode of bubble propagation to increases in the value of $s$ at fixed $\alpha_w$ (see figure \ref{fig:BifTrack}c,d). 

 Figure \ref{fig:BifTrack}a shows the rail profiles used in the model
 when $s=40$ by comparison with the two experimental rail profiles
 based on micrometer measurements.
 The model predicts that for low flow rates the symmetric on-rail state is unstable and it
gains stability via a pitchfork bifurcation at a critical value of
the flow rate that depends on the bubble size, for a fixed channel
geometry. Thus, the path of the pitchfork bifurcation should coincide with
the boundary of the stability tongue found experimentally. Bifurcation-tracking calculations were 
performed to determine the location of this bifurcation 
as a function of bubble size, and numerical stability results are compared with experimental data in figure \ref{fig:BifTrack}b. 
We find broad qualitative, but not quantitative,
agreement. For example, the numerical simulations
systematically underpredict the critical flow rates for stability, but
as the width of the rail increases, the critical flow rates decrease
in both the experiments and the model, as does the relative size of the
bubble at minimum critical flow rate. 
The detailed bifurcation structure underlying on and off-rail propagation modes in the model is
more complex than the results shown in figure \ref{fig:BifTrack}b and will be discussed further in section \ref{sec:biftrack}.

 For the narrower rail, the numerical computations yield a
narrower stability tongue and a smaller minimum value of the critical flow
rate than the experiments. Nonetheless, the bubble diameters
corresponding to the minimum value of the critical flow rate agree to
within 2\% of the numerical value. In order to explore the sensitivity to
the rail geometry, we adjusted the value of $s$ within the range $20 \le s \le 160$, 
while keeping $\alpha_w$ constant -- this is equivalent to maintaining the width of the obstacle between the two points corresponding to half the maximum height (see figure \ref{fig:BifTrack}c). 
The paths of
the pitchfork bifurcation corresponding to these profiles are shown in
figure \ref{fig:BifTrack}d. 
As $s$ increases, the paths converge towards a pointed triangular tongue shape which has a lower cutoff size that is only weakly dependent on flow rate, while the upper bound on drop size increases with flow rate.
The experimental results show a much rounder tongue, reminiscent of the numerical results for smoother profiles.


For the wider rail, one aspect
of the qualitative agreement is lost because, in the model, the
path of the pitchfork bifurcation is such that
larger bubbles remain stable above a critical flow rate. Hence,
according to the model, the system no longer has a large-diameter
cut-off. Nonetheless, the remnants of a tongue-like region remain and
its approximate width is in better agreement with the experimental
data than for the narrow rail.

\begin{figure}
 \includegraphics[width=\columnwidth]{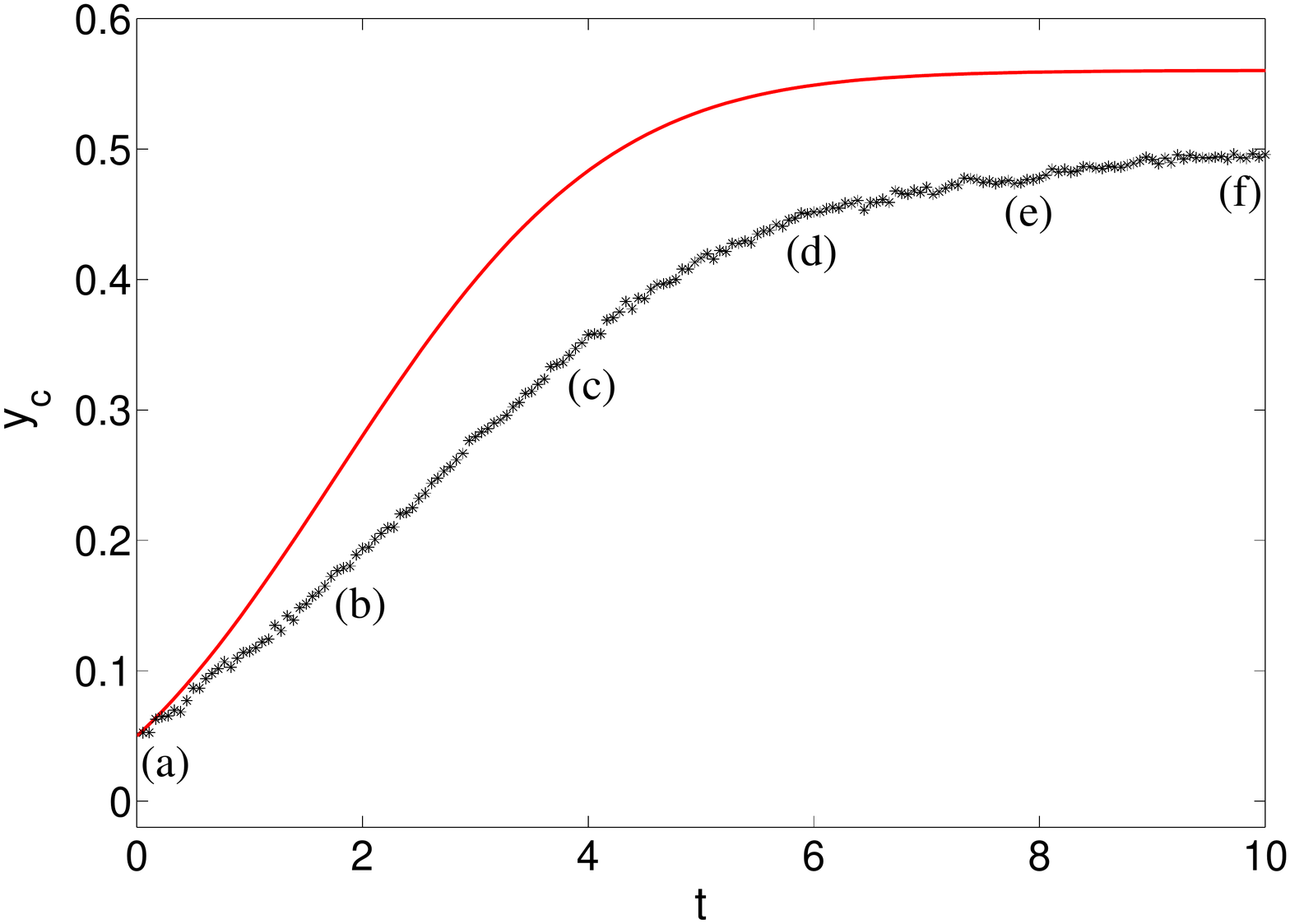}\\
 \ifShapeComparison
 \includegraphics[width=\columnwidth]{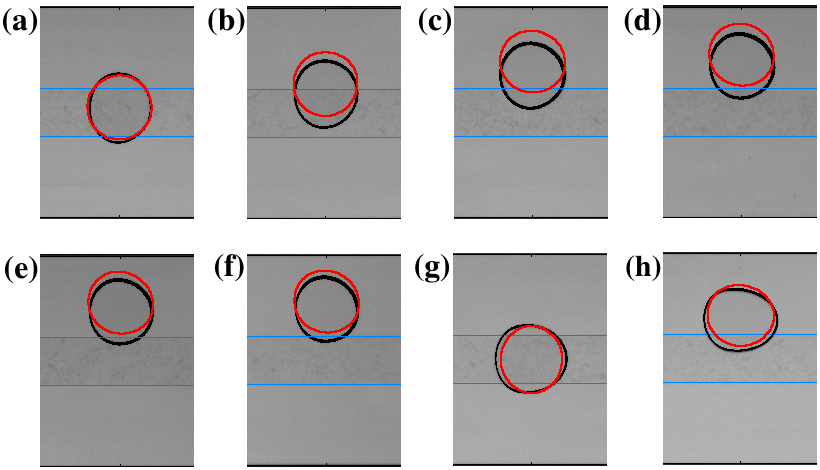}
 \fi
 \caption{Experiments (black) and model predictions (red line) for the evolution of bubble offset $y_c$ as a function of dimensionless time $t$. Here $D=1.33$ and $Q^* = 3 \mlmin$ ($Q=4.0\times 10^{-3}$). 
 \ifShapeComparison
 Snapshots (a)-(f) correspond to times indicated in the first plot. (g) and (h) show steady states for larger flow rates: $Q^* = 19 \mlmin$ ($Q=6.6\times 10^{-3}$) and $Q^* = 13 \mlmin$ ($Q=9.3\times 10^{-3}$) respectively. Sequences of experimental snapshots corresponding to (g) and (h) can be found in figure \ref{fig:Cen_flowrateSchem}.
 \fi
 \label{fig:transient_evolution}  }
\end{figure}

A further comparison between experiment and model can be obtained by studying the transient evolution into the off-rail state. Figure \ref{fig:transient_evolution} shows results for $y_c$ as a function of dimensionless time for a bubble with $D=1.33$ during its migration to an off-rail state at a low flow rate ($Q^* = 3 \mlmin$, $Q=4.0\times 10^{-3}$), and (a)-(f) shows snapshots of bubble shape during this transition. The timescales of bubble motion are in broad agreement, with the predicted migration being faster than the experiments by a factor of approximately 1.6, and the shapes are in reasonable agreement given the evident difference in timescales. The discrepancy in timescales reduces as flow rate is increased for the three cases we have data for, with factors of $1.2$ at $Q=6.6 \times 10^{-3}$ and $1.0$ at $Q=9.3\times 10^{-3}$.  
\ifShapeComparison
The shapes shown in (g) and (h) in figure \ref{fig:transient_evolution} correspond to steady propagation rather than transient migration, and also correspond to (b) and (c) in figure \ref{fig:Cen_flowrateSchem}. 
The main difference between the observed and predicted shapes is that the predicted bubbles have smaller projected area, which will be discussed shortly.
\fi

The numerical results presented in figures \ref{fig:BifTrack} and \ref{fig:transient_evolution} suggest
that the model captures the qualitative features of bubble propagation
in the channel with a centred rail, and thus may be used as a design
guide rather than a predictive tool. The depth-averaged lubrication
model is convenient because it significantly reduces computational
costs compared with a three-dimensional model based on the
Navier--Stokes equations. It also dissociates the effect of the rail on
the bulk viscous flow from its effect on surface tension forces
through cross-sectional curvature variations which are accounted for
in the dynamic boundary condition. This
facilitates numerical investigation into the physical mechanisms of
on-rail bubble propagation because the effect of the channel
topography may be removed from either equation independently, as will be discussed in section
\ref{sec:physical}. However, the quantitative discrepancies between
the depth-averaged model and the experiment are most likely to be due to
three-dimensional effects. Although the channel has an aspect ratio
$\alpha=30$, the width to depth ratio of the confined bubbles is as
low as $7$.  Moreover,  the depth-averaged model does not account for
the liquid films that separate the bubble from the boundaries of the
channel and the rail. For finger propagation, the presence of such films has been found to
 generate
quantitative discrepancies with the model for comparable values of the
non-dimensional flow rate investigated here
\cite{FrancoGomez_etal2016}.

Film thicknesses were estimated by
measuring the increase in mean bubble diameter (relative to the static
bubble diameter)  with increasing flow rate, for a bubble that retains
a constant volume. For $Q=2.50 \times 10^{-2}$ (or $Q^*=19$~ml/min),
which is a typical experimental flow rate in figure
\ref{fig:BifTrack}b,d, and also shown in figure \ref{fig:transient_evolution}g,  the film thickness is approximately four times
larger than the rail height (see Supplementary Material). This suggests that the bubble may not feel the solid rail but an
effective rail profile covered by a fluid film. It also concurs with the
stability results of figure  \ref{fig:BifTrack}b,d, which indicate that wider
rails with gently sloping side walls are required to approximate the
experimental measurements. Although it would be possible to apply an
optimisation procedure to find an effective geometry for the
depth-averaged model that best fits the experimental data,
this does not lead to a predictive model. Hence, it appears that a detailed
simulation of the thin films would be essential to achieve a
quantitative prediction of the stability tongues. The feasibility of
such simulations in the absence of rails where lateral symmetry can be
assumed has recently been demonstrated\cite{Ling2016},
but the simulations remain too computationally expensive to conduct
a detailed bifurcation analysis in reasonable time.
  
\subsection{Bifurcation diagram}
\label{sec:biftrack}

  \begin{figure}
   \includegraphics[width=0.95\columnwidth]{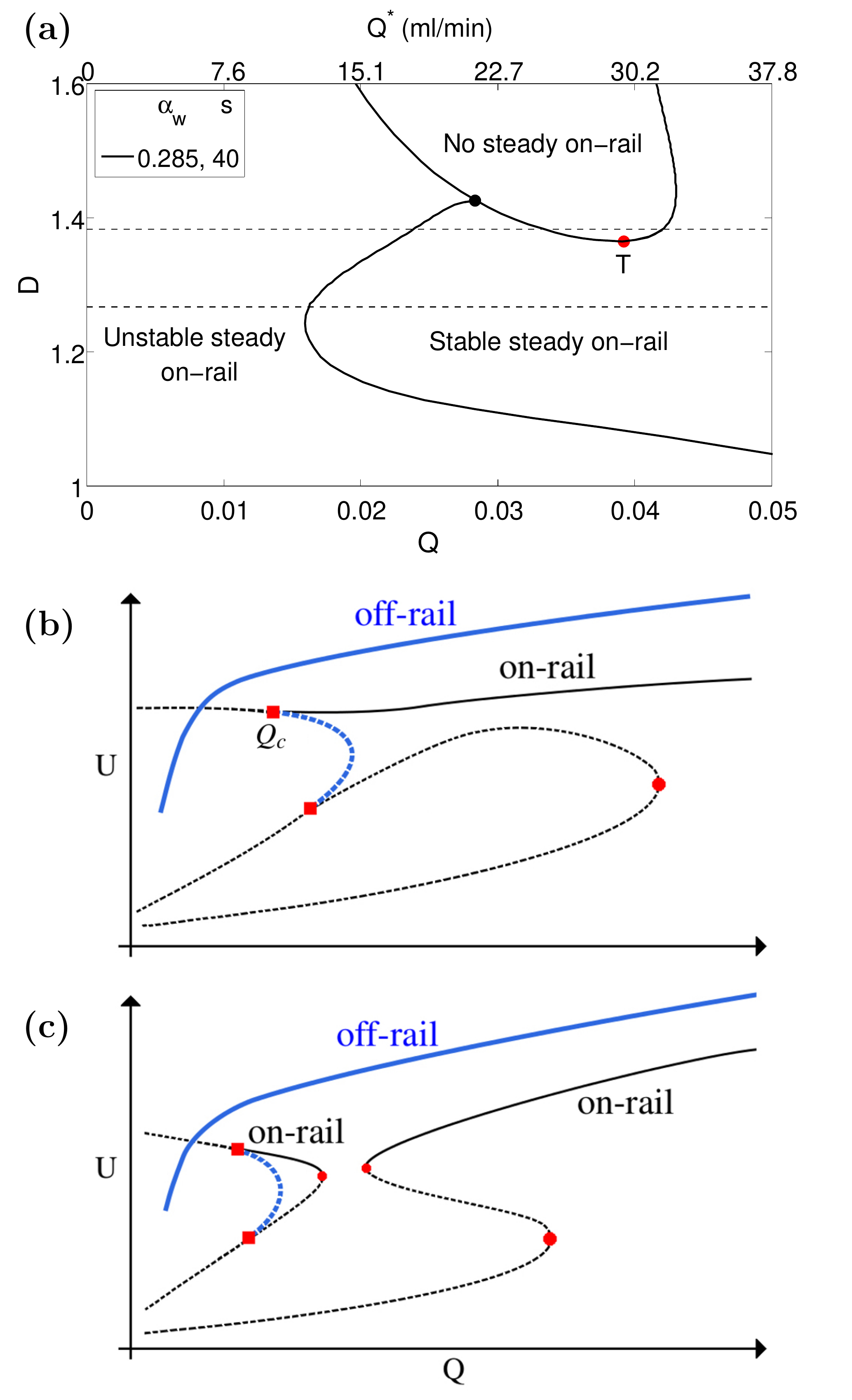}
     \caption{\small (a) Bifurcation diagram for a channel with
       $\alpha=30$, $\alpha_w=0.285$, $\alpha_h=0.024$
       and $s=40$ spanning flow rate and bubble diameter. The path of a pitchfork bifurcation separates regions of unstable and stable steady on-rail propagation. The path of a limit point encloses a region in which there are no steady
       on-rail states. Schematic bifurcation diagrams of the bubble speed as a function of flow rate are shown for two values of the bubble diameter indicated with dashed lines: (b)  $D=1.27$ ($D^*=8.74$ mm) and (c) $D=1.37$
       ($D^*=9.44$ mm). Solid (dashed) lines denote stable (unstable) solutions, while blue (black) lines denote off-rail (on-rail) states, respectively. In (b), the upper \textcolor{red}{$\blacksquare$} denotes the pitchfork bifurcation point that stabilises the on-rail state. Upon the increase of the bubble diameter, the stable on-rail solution branch interacts with the unstable RVB solutions in the lower part of diagram (b), which results in its disconnection in (c) so that there is no simple steady on-rail solution in the region delimited by the upper two saddle node points \textcolor{red}{\Large\bf$\bullet$} in (c), although the unstable RVB solution remains in the lower part of the diagram. The disconnection of the stable on-rail solution branch arises through a transcritical bifurcation at $D \approx$ 1.365 ($Q=0.039$), marked as T in (a).}
       \label{fig:Bifurcations_Speed_highCa}
  \end{figure}

  The stabilisation of the symmetric on-rail propagation modes occurs
 via a symmetry-breaking pitchfork bifurcation, but an obvious
 question is what happens to this bifurcation for small and large
 bubbles. In other words: how is the stability of the on-rail
solutions lost as the bubble size changes? From a dynamical systems
point of view, the easiest scenario would be
that the symmetry-breaking bifurcation rapidly moves to a very high
flow rate as the critical bubble sizes are reached, leading to the
parallel-sided tongue sketched in figure \ref{fig:ExpBistability2}.
In our model we do indeed find this to be the case
for small bubble diameters, but the scenario for
larger bubble sizes is more complex.

Figure \ref{fig:Bifurcations_Speed_highCa}a shows the results of steady bifurcation
tracking calculations extending the results for the narrower rail ($s=40$, $\alpha_h=0.024$ and
$\alpha_w=0.285$) shown in figure \ref{fig:BifTrack}b. We find that there is a region, bounded
by a path of limit points, where the steady
on-rail solution previously investigated is completely absent.
For large bubble diameters the pitchfork bifurcation is annihilated at
this path of limit points at a finite value of the flow rate (black solid circle). In time-dependent
simulations, we find that bubbles in the region in which there are no
steady on-rail solution evolve until they reach a point at which they would be expected
split into multiple bubbles, but bubble pinch-off is not included in
our current model.

The gap in steady on-rail solutions can be further understood by considering
the bubble velocities as functions of flow rate for all computed steady
solutions at a fixed bubble size, see figures \ref{fig:Bifurcations_Speed_highCa}b,c, which
correspond to $D=1.27$ and $D=1.37$, respectively. In
figures \ref{fig:Bifurcations_Speed_highCa}b,c, solid (dashed) lines denote stable (unstable)
propagation modes, and on-rail (off-rail) bubbles are shown with
black (blue) lines, respectively. In both cases, the
off-rail mode of propagation is stable for all values of
$Q$. On-rail (symmetric) bubbles generally travel with lower speeds than the
off-rail (asymmetric) bubbles because they are confined to a shallower channel
over the rail, and thus viscous resistance is relatively increased.
In figure \ref{fig:Bifurcations_Speed_highCa}b, the on-rail bubble is
unstable for flow rates below a critical value $Q_c$ and stable
above. In fact, the loss of stability of the stable on-rail bubble at
$Q_c$,  represented by  a stability tongue as a function of bubble
size  in figure \ref{fig:Bifurcations_Speed_highCa}a, occurs through a
subcritical pitchfork bifurcation at $Q_c$, where two unstable asymmetric
branches emerge (these branches have the same bubble velocity $U$). These
asymmetric solution branches connect via a second pitchfork
bifurcation to another type of symmetric
solution of lower bubble velocity. This symmetric solution is unstable, and exhibits the
characteristically dimpled bubble tips of unstable
Romero--Vanden-Broeck solutions (RVB) uncovered in the context of
Saffman--Taylor fingering in Hele-Shaw channels
\cite{Romero1982,VandenBroeck1983}. 
As the bubble
size increases from $D=1.27$ to $D=1.37$, interaction between the stable
on-rail solution branch and the unstable RVB solution branch leads to
a transcritical exchange of stability at an intermediate value of
$D$. A further increase in bubble size towards $D=1.37$ disconnects the
transcritical bifurcation through the emergence of two limit points,
which bound an interval of intermediate values of $Q$ where stable
on-rail bubbles are not found, as illustrated in figure
\ref{fig:Bifurcations_Speed_highCa}c. The variation of these limit
points with increasing bubble size form the stability boundary in
figure \ref{fig:Bifurcations_Speed_highCa}a that encloses the `no
steady on-rail' region. As the bubble size increases further the
two pitchforks move closer together and annihilate each other as
they interact with the limit point.

\subsection{Physical mechanism of on-rail propagation}
\label{sec:physical}

 We now investigate the physical
 mechanisms that enable on-rail bubble propagation. For consistency
 with the numerical results, we shall describe the mechanisms via the
 depth-averaged model described in section \ref{nummod}. In this
 framework, the deformation of the bubble is driven by the kinematic
 condition, equation (\ref{VeloCont}), and in the lab frame the normal
 velocity is given by the term $-b^{2} {\bf\hat n}\cdot \bm{\nabla} p$.
 Hence the behaviour of the bubble can be inferred from the local
 pressure gradients and channel depth. Note that the depth-averaged velocity profile that results from this model in the absence of obstacles and bubbles is uniform and not parabolic.

  \begin{figure}
 \vspace{-1em}
\includegraphics[width=0.90\columnwidth]{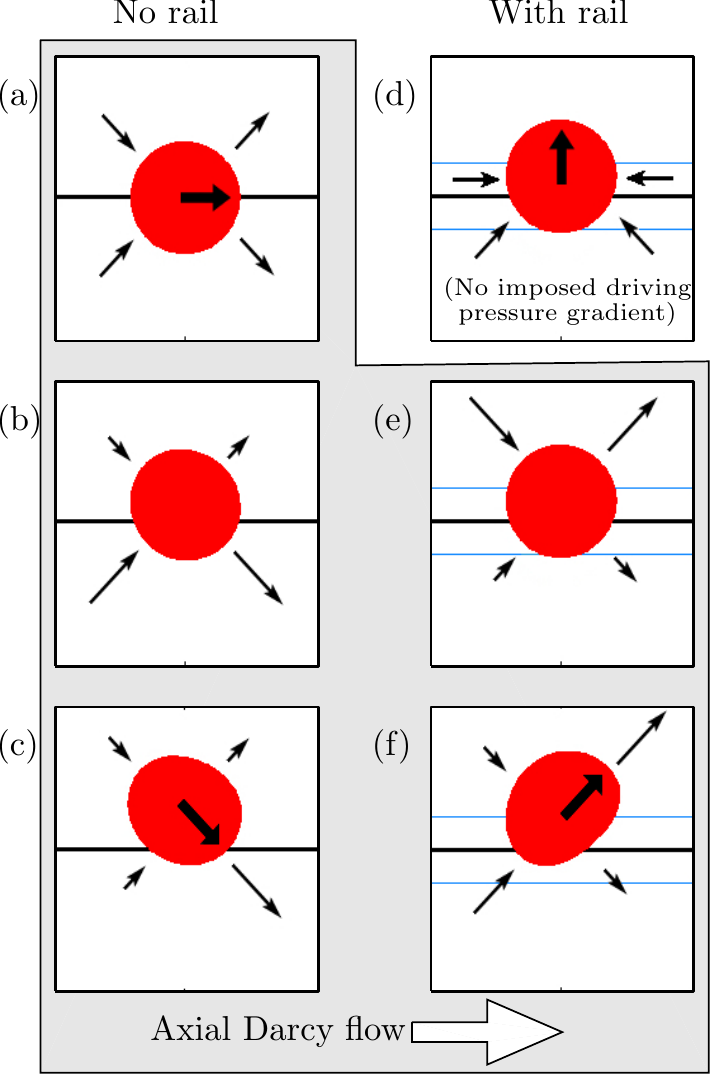}
\caption{\small Schematic diagram of migration mechanisms in
  an unoccluded Hele-Shaw cell (a)--(c) and in presence of a rail
  (d)--(f). The system is subject to an imposed pressure gradient driving a net flow from left to right in all cases except (d).
Small arrows are  normal velocity vectors and the large
  arrow indicates net direction of travel of the bubble. The
  centreline is shown as a solid line and in (d)--(f) the limits of
  the centred rail are shown as dotted lines. (a) Symmetric
  velocity field for a centred bubble. (b) Displacing the bubble reduces
  the normal pressure gradients and hence the normal velocities on the
  side of the bubble nearest the side wall, which leads to a net
  inclination of the bubble shown in (c). The surface-tension-induced
  pressure jump serves to enhance the pressure gradients, and hence
  normal velocities, near the tips, driving propagation in the
  direction of inclination: back towards the centreline. 
  If there is no imposed driving pressure gradient (d), the presence of the
  rail leads to an increased cross-sectional curvature, which locally
  lowers the fluid pressure leading to positive pressure gradients
  away from the bubble and hence normal velocities over the rail directed towards
  the bubble, as shown. These drive net migration into the side
  channel. (e) Displacing the bubble when the rail is present leads to a net
  increase in normal velocities on the side of the bubble nearer the
  wall owing to the presence of the deeper side channels, provided
  that the increase in depth is sufficient to overcome the
  reduction in normal pressure gradient described in (b). The higher
  velocities nearer the wall lead to a net inclination of the bubble
  towards the side channel shown in (f), in the opposite direction
  to that in the unoccluded case (c). Thus, in this case when the
  bubble migrates in the direction of inclination it moves
  into the side channel.}
 \label{fig:MechanismSchematic}
  \end{figure}

 In unoccluded Hele-Shaw cells an
 interaction between viscous and capillary effects
 underlies the dynamic mechanism that acts to return off-centre bubbles
 to the centreline. An initially circular bubble (when viewed from
 above) travelling along the
 centreline of an unoccluded Hele-Shaw call 
 is driven by pressure and velocity fields that are symmetric about
 the centreline, which induce a symmetric
 deformation, see figure \ref{fig:MechanismSchematic}a.
 For small bubbles, this results in a slight decrease
in  the in-plane curvature (when viewed from above)
 at the front of the bubble and an increase 
 in the in-plane curvature at the rear.
 If the bubble is displaced from the centreline, the magnitude of the 
 pressure gradients normal to the bubble will decrease
 on the side of the bubble nearest the side walls where $\partial
 p/\partial y = 0$, see figure \ref{fig:MechanismSchematic}b. The ensuing 
 asymmetric velocity field will cause the bubble to
 incline relative to the centreline such that the point of maximum
 in-plane curvature at the rear moves towards the sidewall and the point of
 minimum in-plane curvature at the front moves towards the centreline,
 \textsl{i.\@e.\@}
 the bubble ``points'' towards the centreline, see figure
 \ref{fig:MechanismSchematic}c.
 The surface-tension-induced pressure jump over the 
 interface means that the fluid pressure
 adjacent to the bubble will be lowest at point of 
 maximum curvature and highest at the point of minimum curvature, 
 which increases the magnitudes of the 
 local pressure gradients and hence drives the bubble
 in the direction of inclination,
 eventually returning the bubble to the centreline. 
 
 Outside the depth-averaged framework the no-slip boundary condition on the channel walls leads to a non-uniform velocity profile with the fastest flow along the channel centreline. The velocity profile causes a bubble displaced from the centreline to incline in the same direction as a displaced bubble in the depth-averaged model and subsequently the same mechanism
drives the conventional migration towards the
centreline for Stokes flows in unoccluded channels
 \cite{Richardson1973,ChanLeal_etal1979}. The
 propagation of bubbles in the direction of inclination has also
 recently been implicated as a key mechanism in Leonardo's paradox: the
 onset of zigzag or helical trajectories for rising bubbles in still
 liquid\cite{CanoLozano2016}.
 These migration effects do not occur
 for rigid particles because they do not change their shape in
 response to the velocity field. We remark that
 the interaction between deformation and
 capillary-induced pressure jump is essential for the mechanism to
 operate. 
 
\begin{figure}
    \includegraphics[width=0.85\columnwidth]{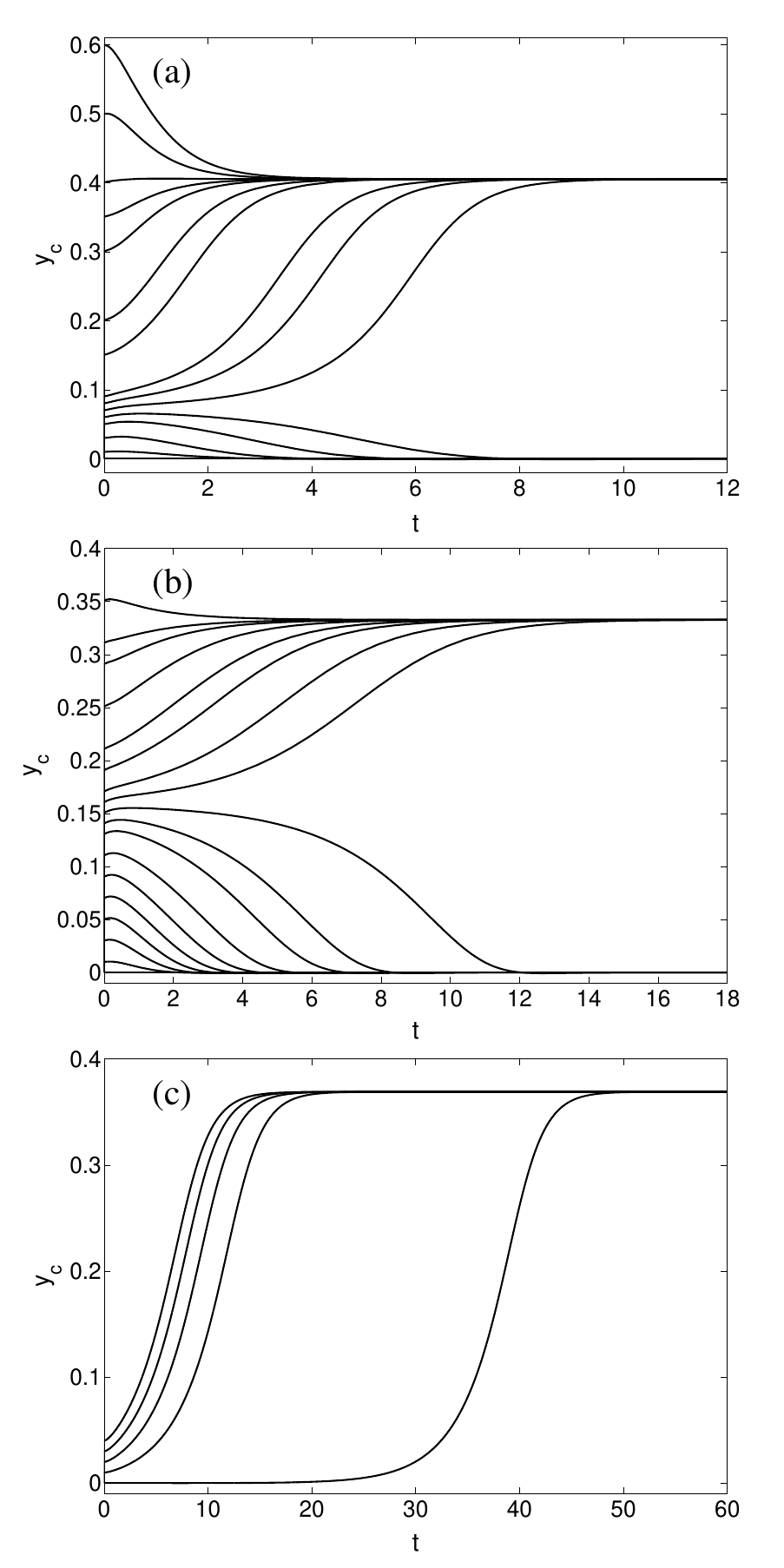}
\vspace{-1em}
     \caption{\small Bubble centroid position (y-coordinate)  as a function of nondimensional time from time-dependent numerical simulations. The channel parameters parameters are $\alpha=30$, $\alpha_w=0.285$ (narrow rail), $\alpha_h=0.024$ and $s=40$. The bubble has diameter $D=1.27$ ($D^*=8.74$ mm) and propagates with a fixed flow rate of $Q=2.60 \times 10^{-2}$ ($Q^*=20$ ml/min), but note that we conduct this numerical experiment at fixed 
pressure gradient. (a) The prescribed channel depth variation $b(y)$ is included in both the bulk equations (\ref{LaplaceEq},\ref{VeloCont}) and the dynamic boundary condition (\ref{PressureJump}). The bubble remains on-rail for initial centroid positions $y_c\leq0.06$ ($6$ \%) and the steady-state off-rail position is $y_c=0.405$. (b) The prescribed depth variation is  retained in the dynamic boundary condition but removed from the bulk equations by setting $b(y)=1$. The bubble remains on-rail for initial centroid positions $y_c\leq0.15$ ($15$ \%) and the steady-state off-rail position is $y_c=0.333$. (c)  The prescribed depth variation is retained in the bulk equations but removed from the dynamic boundary condition by setting $b(y)=1$. The bubble migrates to an off-rail steady state with $y_c=0.369$  for all initial centroid positions.}\label{fig:PertCentNumStepsEffects}
  \end{figure}

 The introduction of
 a rail alters both capillary forces via changes in cross-sectional
 curvature and also viscous forces via changes to the bulk resistance
 of the channel. Thus, we should expect the restoring mechanism just
 described to be disrupted for sufficiently large rails. 
 In order to investigate the physical mechanisms leading to the migration of bubbles in the presence of the rail, we conducted a number of time-dependent simulations, selecting which terms are affected by the rail.
 For this numerical experiment, we chose the narrow rail, a fixed bubble size ($D=1.27$), a fixed flow rate ($Q = 2.60\times 10^{-2}$), and a range of initial
centroid positions. The results are shown in figure
\ref{fig:PertCentNumStepsEffects}, where the centroid position $y_c$
of the bubble is shown as a function of time. 

In figure
\ref{fig:PertCentNumStepsEffects}a, cross-sectional depth
variations in the channel are retained both in the bulk equations
(\ref{LaplaceEq}, \ref{VeloCont}) and in the dynamic boundary
condition (\ref{PressureJump}). Only a nearly symmetric 
initial bubble placement ($|y_{ci}|
\le 0.06$) results in stable on-rail bubble propagation; in other
words, the basin of attraction of this stable solution is relatively
small. As expected, the duration
of transient migration is greatest for initial conditions near the
basin boundary value of $y_{ci}$ separating stable on and off-rail
bubble propagation. In figure \ref{fig:PertCentNumStepsEffects}b, the
effect of depth variation is removed in the bulk equations (\ref{LaplaceEq},\ref{VeloCont})
by setting $b(y)=1$, but the depth-profile is
retained in the dynamic boundary condition
(\ref{PressureJump}). Removal of the variations in viscous resistance
more than doubles the range of initial centroid positions ($|y_{ci}|
\le 0.15$) for which stable on-rail propagation is achieved. Hence,
depth variations in the bulk enhances destabilisation of on-rail
states. Finally, in figure
\ref{fig:PertCentNumStepsEffects}c, the effect of cross-sectional
depth variations is retained in the bulk but not in the
dynamic boundary condition by setting $b(y)=1$ in
eq. (\ref{PressureJump}). In this case,
all initial bubble positions lead to stable off-rail propagation.
This demonstrates that the
cross-sectional curvature variation introduced by the depth profile is
essential to promote stable on-rail propagation for the system at these particular parameter values; 
a rather surprising result given that the very same curvature
variation is responsible for destabilisation of the static bubble. 
  
Thus, the variation of viscous resistance caused by
introduction of the rail in the absence of changes in cross-sectional curvature may lead to destabilisation of the
on-rail propagation mode and eventual migration towards the off-rail
propagation modes, despite the fact that the on-rail propagation mode is not significantly different from that shown in figure \ref{fig:MechanismSchematic}a. The non-uniform
viscous resistance means that for a constant axial pressure gradient
the velocity is faster in the
off-rail regions, see figure \ref{fig:MechanismSchematic}e.
If the bubble is displaced
from the centreline and partially enters the off-rail region
then the resulting 
speed differential over the interface will cause it to
incline relative to the centreline so that it appears to be directed
into the off-rail region, see figure \ref{fig:MechanismSchematic}f.
The surface-tension-induced pressure jumps over the interface will
again cause the bubble to migrate in the direction of inclination,
which is now towards the off-rail region. As the bubble
moves laterally more of the interface enters 
the off-rail region and travels at the
higher speed, meaning that the bubble will become less inclined. 
Thus, the lateral
propagation velocity decreases until the stable off-rail propagation
mode is reached. If the viscous resistance remains uniform then this
destabilisation mechanism is absent, which explains why
the basin of attraction of the on-rail state increases in this case.
For the parameters chosen in figure \ref{fig:PertCentNumStepsEffects} we find that by reducing the rail height to less than one third of its former value, the standard centering mechanism is able to overcome these effects of  bulk variation, and the on-rail state is again stabilised.

 In the absence of flow, the
increased cross-sectional curvature caused by confinement leads to locally lower pressures near the regions of the interface over the rail. The resulting pressure gradients drive flows that destabilise a bubble displaced from the centreline, which moves into an off-rail position, see figure
\ref{fig:MechanismSchematic}d.
If there is no variation in the
viscous resistance, however, then in the presence of flow,
as in the unoccluded channel, the bubble
inclines so that its tip is nearer
the centreline, which acts to drive the bubble towards an on-rail
position. Thus, the capillary-driven and
viscous-pressure-driven flows act in opposite directions near the 
downstream end of the bubble; compare the arrows on the right-hand-side of the bubbles in figures \ref{fig:MechanismSchematic}c and \ref{fig:MechanismSchematic}d. Hence, for
sufficiently large flow rates, the on-rail position is always
stabilised when there is no variation in viscous resistance.

 When variations in depth are included in both bulk equations and
interfacial boundary conditions then the final direction of
inclination of the bubble caused by the interplay between
viscous and capillary effects, and hence stability of the on-rail
state, depends crucially on the bubble size and
position, which gives rise to the observed complex behaviour for
which we have no simpler predictive model than solving the full 
set of two-dimensional governing equations.
  
\begin{figure}
    \begin{center}
       \includegraphics[width=\columnwidth]{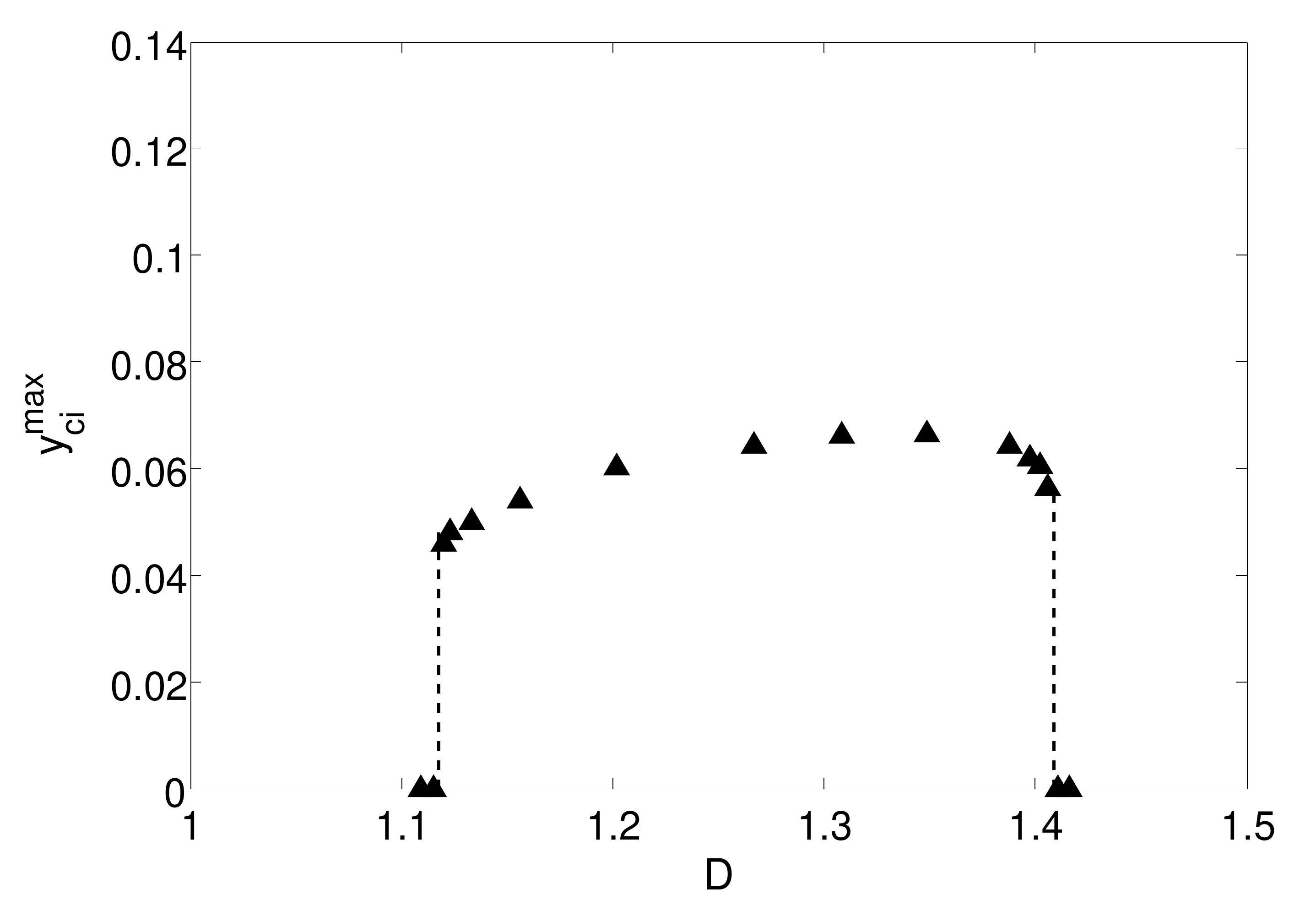}
    \end{center}
    \vspace{-0.5cm}
    \caption{\small Maximum initial centroid position allowing a bubble to return to a stable on-rail propagation mode as a function of bubble diameter in the numerical simulations. The flow rate and channel parameters are the same as in figure \ref{fig:PertCentNumStepsEffects}.} \label{fig:PressJumpCentallBubbs}
  \end{figure}

The time-dependent numerical simulations shown in figure
\ref{fig:PertCentNumStepsEffects}a were repeated for a wide range of
bubble sizes. Figure \ref{fig:PressJumpCentallBubbs} shows the largest
initial off-centre centroid position $y_{ci}^{\mathrm max}$ for which
stable on-rail propagation occurs as a function of bubble size, within the range of bubble sizes
for which stable on-rail propagation could be achieved. The variation
of $y_{ci}^{\mathrm max}$ is small and there is a sudden drop in
$y_{ci}^{\mathrm max}$ at $D=1.12$ and $D=1.41$ which are the
smallest and largest bubble sizes that allow stable on-rail
propagation, respectively. This suggests that the sensitivity to
positioning perturbations remains essentially constant within the
stability tongues previously shown in  figure {\ref{fig:BifTrack}b}. 

\section{Conclusions}  \label{conc}

The transport of bubbles, capsules and droplets by driving
carrier fluid within a
confined geometry is a universal process in industry and
nature. The channel geometry influences
the possible propagation modes and for a
large-aspect-ratio rectangular channel, or Hele-Shaw cell, 
the solution structure is particularly rich
\cite{Tanveer1987,Kopf-sillHomsy1988}. A
panoply of different propagation modes has been uncovered, but for
Hele-Shaw channels without rails or grooves, the vast majority of
these modes are unstable and the only observed propagation mode
is centred and symmetric about the channel's centreline.

 In the present paper, we find that introduction of a rail on the base
 of the channel introduces stable asymmetric,
off-rail propagation modes for all flow rates and bubble sizes
considered. In contrast, the centred,
on-rail propagation modes are only stable for a limited range of
bubble sizes and above a minimum flow rate $Q_{cm}$. Thus, bubbles
within a narrow size band may be segregated
from others by fixing the flow rate at a value marginally
larger than $Q_{cm}$.  Under these conditions, the propagation of a
train of bubbles of different sizes, which are sufficiently separated
to avoid bubble interactions, will result in only those 
bubbles within the band remaining on the rail after reaching a steady
propagation state. Both smaller and larger bubbles will migrate into the
side channels allowing their removal from the main stream as
illustrated schematically in figure \ref{sorting}. The experiments
presented in this paper were performed with quasi-static initial
conditions, but in practice, bubbles propagating steadily in a channel
of rectangular cross-section would be centred
\cite{ChanLeal_etal1979,Stan_etal2011}, and they could thus encounter
the rail at a small distance downstream of the inlet of the channel,
where segregation would occur.

 In contrast to transport along grooves \cite{Abbyad2011}, which
relies on stabilisation of the capillary static solution, the
present mechanism is dynamic, requiring sufficiently large viscous
restoring forces to compensate for destabilising capillary-driven
flows. The restoring mechanism is subtle, however, because both capillary and viscous forces can act as destabilising or stabilising in different regimes. The mechanism relies on an interplay 
between the viscous pressure gradient and
surface-tension-induced capillary pressure drop, which leads to a
critical flow rate for stable on-rail propagation that is not a
monotonic function of bubble size.

 We have restricted attention to bubbles of a comparable size to the
rail width. As the bubbles increase in size and eventually interact with the
sidewalls of the channel then they can no longer be segregated using
this mechanism. In fact, large bubbles  in Hele-Shaw cells
under the introduction of a rail exhibit different but equally complex behaviour, similar to that of air fingers \cite{FrancoGomez_etal2016}. Their dynamics are currently under investigation.

\section{Acknowledgements}  

A.F-G. was funded by CONICYT. We thank Pallav Kant for performing the profilometry measurements of the thin-film rail, and Lucie Duclou\'e for helpful discussions. This work was supported
by the Leverhulme Trust (Grant RPG-2014-081) and EPSRC (Grant EP/P026044/1).

\footnotesize{ 
\providecommand*{\mcitethebibliography}{\thebibliography}
\csname @ifundefined\endcsname{endmcitethebibliography}
{\let\endmcitethebibliography\endthebibliography}{}

\bibliographystyle{rsc} }

\end{document}

\end{document}